# Spin liquid and ferroelectricity close to a quantum critical point in PbCuTe$_2$O$_6$

Christian Thurn[1*], Paul Eibisch[1], Arif Ata[1], Maximilian Winkler[2], Peter Lunkenheimer[2], István Kézsmárki[2], Ulrich Tutsch[1], Yohei Saito[1], Steffi Hartmann[1], Jan Zimmermann[1], Abanoub R. N. Hanna[3,4], A. T. M. Nazmul Islam[4], Shravani Chillal[4], Bella Lake[3,4], Bernd Wolf[1], Michael Lang[1]

[1]Physikalisches Institut, J.W. Goethe-Universität Frankfurt(M), 60438 Frankfurt, Max-von-Laue-Str.1, Germany

[2]Experimentalphysik V, Zentrum für Elektronische Korrelationen und Magnetismus, Universität Augsburg, Universitätsstr. 1, 86159 Augsburg, Germany

[3]Institut für Festkörperforschung, Technische Universität Berlin, Hardenbergstr. 36, 10623 Berlin, Germany

[4]Helmholtz-Zentrum Berlin für Materialien und Energie, Hahn-Meitner Platz 1, 14109 Berlin, Germany

*e-mail: Thurn@Physik.uni-frankfurt.de

**Geometrical frustration among interacting spins combined with strong quantum fluctuations destabilize long-range magnetic order in favour of more exotic states such as spin liquids. By following this guiding principle, a number of spin liquid candidate systems were identified in quasi-two-dimensional (quasi-2D) systems. For 3D, however, the situation is less favourable as quantum fluctuations are reduced and competing states become more relevant. Here we report a comprehensive study of thermodynamic, magnetic and dielectric properties on single crystalline and pressed-powder samples of $PbCuTe_2O_6$, a candidate material for a 3D frustrated quantum spin liquid featuring a hyperkagome lattice. Whereas the low-temperature properties of the powder samples are consistent with the recently proposed quantum spin liquid state, an even more exotic behaviour is revealed for the single crystals. These crystals show ferroelectric order at $T_{FE} \approx 1$ K, accompanied by strong lattice distortions, and a modified magnetic response – still consistent with a quantum spin liquid – but with clear indications for quantum critical behaviour.**

## Introduction

Competing interactions combined with strong quantum fluctuations are considered a major guiding principle for the realization of a quantum spin liquid. This long sought state of matter is characterized by exciting properties such as macroscopic entanglement and fractionalized excitations, see refs. [1-5] for recent reviews. As quantum fluctuations are enhanced for low spin values and low lattice coordination, most efforts have been devoted to low-dimensional (low-D) quantum spin ($S = 1/2$) antiferromagnets, with prominent examples including the 2D kagome system Herbertsmithite[6,7] and some layered triangular-lattice charge-transfer salts[8-10]. For 3D lattices, the search has focused on materials where the spins reside on a pyrochlore[11,12] or hyperkagome lattice[13], a 3D network of corner-sharing triangles. As in 3D competing ground states are expected to become more relevant, perturbations from purely Heisenberg spin scenarios due to, e.g., Dzyaloshinskii-Moriya interaction[14,15] or spin-orbit coupling[16] gain in importance and may eventually govern the materials' ground state[1]. This raises the general question about the stability of a quantum spin liquid phase in 3D. In fact, for $Na_4Ir_3O_8$, a 3D effective $S = 1/2$ spin liquid candidate system with a hyperkagome lattice[13], evidence was reported for a nearby quantum critical point[17] – a $T = 0$ instability to some nearby (possibly magnetically) ordered state. More recently, $PbCuTe_2O_6$ has attracted considerable interest as another 3D quantum spin liquid candidate system featuring a highly-connected hyperkagome lattice[18-20], where $S = 1/2$ $Cu^{2+}$ spins are coupled by isotropic antiferromagnetic interactions. According to magnetic studies, mainly on pressed-powder samples[18-20], this system lacks long-range magnetic order down to 0.02 K[19] and shows diffuse continua in the magnetic spectrum[20] consistent with fractional spinon excitations. Puzzling issues relate to the appearance of small anomalies in the powder samples around 1 K of unknown origin[18-20] and signs of a phase transition around this temperature in first-generation single crystals[20].

In the present work, we present a comprehensive study of thermodynamic, magnetic and dielectric properties of $PbCuTe_2O_6$, with special focus lying on its low-temperature state. To this end, we have investigated several single crystals and compare the results with data on pressed-powder samples. The salient results of our study are (i) the observation of a phase transition in single crystalline material around 1 K into a ferroelectrically-ordered state, which is accompanied by strong lattice distortions. (ii) This phase transition and the state at $T \leq 1$ K are characterized by a finite magnetic susceptibility without any indication for long-range magnetic order consistent with a gapless quantum spin liquid state. (iii) The bulk ferroelectric transition and its accompanying lattice distortions are absent in the powder samples. (iv) For these powder samples, the ratio $\alpha/C_e$, with $\alpha$ the thermal expansion coefficient and $C_e$ the specific heat, probing the thermal Grüneisen parameter $\Gamma_p$, is practically independent of temperature for $T \leq 1.6$ K. In contrast, a corresponding uniaxial Grüneisen parameter for the single crystals shows a strong increase for $T \rightarrow 0$, potentially indicating quantum critical behaviour. In fact, clear evidence for a nearby field-sensitive quantum critical point is observed in the electronic magnetic Grüneisen parameter $\Gamma_{B,e}$, yielding a divergence for $T \rightarrow 0$.

## Results

**Material, crystal structure and magnetic interactions**

$PbCuTe_2O_6$ crystallizes in a non-centrosymmetric cubic structure with space group $P4_132$ (No. 213), see Supplementary Fig. 1 and Note 1. According to density functional theory calculations[20] based on room-temperature structural data, the magnetic lattice can be described by isolated equilateral $S = 1/2$ triangles with nearest-neighbour interaction $J_1 = 1.13$ meV, which are coupled via the second-nearest neighbour interaction $J_2 = 1.07$ meV into a hyperkagome lattice. The weaker

third- and fourth-nearest neighbour interactions $J_3$ = 0.59 meV and $J_4$ = 0.12 meV couple the spins into chains. Another special feature of PbCuTe$_2$O$_6$, which has been largely ignored until now, relates to its dielectric degrees of freedom[21]. The material contains polar building blocks originating from the free electron pairs (lone pairs) of the Te$^{4+}$ ions in the oxotellurate tetrahedrons and the asymmetrically coordinated Pb$^{2+}$ ions. These characteristics together with the non-centrosymmetric structure imply the possibility of ferroelectric order interacting with the strongly frustrated quantum spin system. In fact, a magneto-dielectric effect was observed in SrCuTe$_2$O$_6$[22] which is isostructural to PbCuTe$_2$O$_6$ but features a magnetic network with predominantly 1D character.

Measurements were performed on single crystals (sc) from 4 different batches grown by utilizing two different techniques (see Methods and ref. 23), and pressed-powder (pd) samples (see Methods) from the same batch prepared as described in ref. 20. In what follows, the samples will be specified by giving their batch number followed by an alphabetic character to distinguish different samples from the same batch.

**Thermodynamic properties**

Figure 1 shows the results of the electronic specific heat, $C_e$, measured on single crystal sc #5(b). The data reveal a pronounced λ-shape phase transition anomaly around 1 K, signalling a second-order phase transition, on top of a broad maximum. The figure also includes data for a pressed-powder sample pd #I(b) showing a distinctly different behaviour. Here we find the broad maximum in $C_e(T)$ centred around 1.4 K, followed by a smooth reduction upon further cooling similar to previous reports on powder material[18-20]. As also revealed in these studies, we observe signatures for a tiny feature around 1 K and a $C_e(T) \propto T^n$ dependence with $n \approx 2$ at low temperatures $T \leq 0.4$ K

More insight into the nature of the phase transition in the single crystalline material can be obtained by studying the elastic properties via thermal expansion measurements, cf. Fig. 2. In the inset of Fig. 2, we show data of the relative length changes, $\Delta L_i(T)/L_i$, for sc #1(a) measured along two different directions. Based on structure determination at room temperature, we refer to these directions as $i$ = [100] and [1-10]. For the data at higher temperatures down to about 1.7 K, we find a smooth isotropic reduction of $\Delta L_i(T)/L_i$ with decreasing temperature, consistent with a cubic structure. However, on further cooling the thermal contraction becomes increasingly stronger and develops a pronounced anisotropy, indicating deviations from cubic symmetry at low temperatures. At around 1 K we find a sharp break in the slope in $\Delta L_i(T)/L_i$ along both directions, consistent with a second-order phase transition. We stress that the evolution of a non-isotropic lattice strain from a cubic high-temperature state implies the formation of structural domains. As a result, the $\Delta L_i(T)/L_i$ data for $T \leq 1.7$ K in Fig. 2 could potentially be affected by the material's domain structure. In general, the formation of domains can be influenced by the application of uniaxial pressure to the crystal in its high-temperature phase. This is actually the case in our thermal expansion measurements along the measuring direction, where uniaxial pressure of typically 0.01 – 5 MPa is applied, depending on the sample geometry and the chosen starting capacitance[24], see also Methods and Supplementary Note 3 . In the main panel of Fig. 2 we show the coefficient of thermal expansion $\alpha_i(T) = 1/L_i \cdot d(\Delta L_i(T))/dT$ for single crystal sc #5(c) along the [110] direction. Since in this experiment a rather high uniaxial pressure of about (6.5 ± 1.3) MPa was realized, we believe that these $\alpha_{[110]}$ data represent a preferential domain orientation coming close to a mono-domain structure, see Fig. 6 in the Supplementary. Therefore, we confine the discussion of the temperature dependence to these $\alpha_{[110]}$ data. Upon cooling, $\alpha_{[110]}$ shows an extraordinarily strong increase followed by a sharp negative phase transition anomaly slightly below 1 K, and a second maximum around 0.5 K. However, a comparison with the results of $C_e(T)$ for the single crystal sc #5(b) in Fig. 1, showing a maximum around 1 K, suggests another interpretation: there is only a single huge maximum in $\alpha_{[110]}$ centred slightly below 1 K to which a negative λ-shape phase transition anomaly around 1 K is superimposed. The somewhat larger width of the phase transition anomaly in $\alpha_{[110]}(T)$ as compared to $C_e(T)$, may indicate some remaining domain misalignment in the thermal expansion experiment, cf. the narrowing of the phase transition anomaly in $\alpha_{[110]}$ on increasing the uniaxial pressure in Fig. 6 of the Supplementary. Figure 2 also shows thermal expansion data taken on the pressed-powder sample pd #I(a). Whereas the $\Delta L(T)/L$ data for the powder match with the data for the single crystal for $T > 1.6$ K, (see inset of Fig. 2) they deviate at lower temperatures: instead of the strong contraction observed for the single crystal on approaching the phase transition, the powder sample shows a rather smooth behaviour with $\Delta L(T)/L$ gradually levelling off upon cooling. The corresponding $\alpha(T)$ data for the powder sample shown in the main panel of Fig. 2 reveal a broad maximum around 1.4 K followed by a smooth variation $\alpha(T) \propto T^n$ with $n \approx 2$ for $T \leq 0.4$ K.

**Dielectric constant and polarization**

Given the material's non-centrosymmetric structure and its polar building blocks, comprising two subsystems of stereochemically active lone pairs associated with the $Te^{4+}$ and $Pb^{2+}$ ions, the observation of a lattice distortion suggests an involvement of the electric degrees of freedom. To probe the dielectric response, measurements of the dielectric constant were performed on single crystals #1(a), #4 and #5(b) for temperatures $T \leq 1.3$ K. In Fig. 3 we show exemplarily the results of the normalized dielectric constant $\varepsilon'$ for sc #5(b). The data reveal an enhanced background dielectric constant of $\varepsilon'_b \approx 18$ in the temperature range investigated and a well-pronounced peak centred at 0.97 K, signalling a ferroelectric transition at $T_{FE} = 0.97$ K. This assignment is further corroborated by a Curie-Weiss-like behaviour on approaching the maximum in $\varepsilon'$ from both sides, see inset (a) of Fig. 3. In the inset (b) of Fig. 3 we show the frequency dependence of the dielectric anomaly around 1 K. The measurements, covering approximately two decades in frequency, reveal a distinct suppression of the peak with increasing frequency whereas the position of the peak remains essentially unchanged. This behaviour is typical for an order-disorder-type ferroelectric transition[26,28] where electric dipoles that are disordered at high temperature order with a net overall polarization below $T_{FE}$. In contrast, for the pressed-powder sample pd #I(b) we find only a tiny anomaly in $\varepsilon'$ around 1 K, see Fig. 3. This observation together with the smooth variation of $\Delta L(T)/L$ for $T \leq 1.6$ K for the powder sample, reflecting the absence of an anomalous lattice distortion as revealed for the single crystals, indicate that there is no bulk ferroelectric transition in the pressed-powder sample.

To further corroborate the ferroelectric nature of the transition detected at about 1 K, we have investigated the polarization $P$ of single-crystalline PbCuTe$_2$O$_6$ (sc #1(a), **E** || [110]). Figure 4 shows $P(T)$ as measured upon heating after poling the sample with different electrical fields during a preceding cooling run as noted in the figure. Without poling field, no significant polarization was detected, consistent with a multi-domain polar state with even distribution of polarization directions. Prepoling with positive fields between 0.1 and 2.3 kV/cm reveals a successively increasing low-temperature polarization, vanishing above $T_{FE} \approx 1$ K. The observed gradual increase of $P(T)$ below $T_{FE}$ is in accordance with a second-order nature of the polar phase transition consistent with results of the specific heat and thermal expansion (Figs. 1 and 2). Between 1.2 and 2.3 kV/cm, $P$ exhibits only weak variation with field, pointing to a mono-domain state with saturation polarization of the order of 5 $\mu$C/m$^2$. As revealed by the lowest curve in Fig. 4, a negative field of -2.3 kV/cm results in a negative polarization, reflecting a polarization into the opposite direction, as expected for ferroelectrics. The absolute value of $P$, reached for -2.3 kV/cm is somewhat reduced compared to +2.3 kV/cm which may arise from strains within the crystal, slightly favouring one polarization direction.

Another defining property of ferroelectrics is the switchability of the polarization by an electrical field at temperatures below the transition. For the present material, this is demonstrated in the inset of Fig. 4. In this experiment, the sample was first polarized by cooling it to 0.9 K under negative electrical field of -2.3 kV/cm and then applying a positive field of 4.7 kV/cm. The subsequent heating run without field revealed positive polarization, unequivocally proving the in-situ switching of the polarization within the ferroelectric state.

It should be noted that the detected saturation polarization $P_s$ of PbCuTe$_2$O$_6$ of about 5 µC/m$^2$ (= 0.5 nC/cm$^2$) is rather small compared, e.g., to the well-established lone-pair ferroelectric BiFeO$_3$[29] or to classical displacive ferroelectrics like BaTiO$_3$[30], revealing $P_s$ values of several tens of µC/cm$^2$. This is consistent with the rather small amplitude of the anomaly observed in the dielectric constant at the transition (Fig. 3). Small values of $\varepsilon'(T)$ and $P_s$ are, e.g., also found in some improper ferroelectrics like TbMnO$_3$[31]. However, it is conceivable that $P_s$ is higher for field directions other than the presently used **E** ∥ [110] geometry. To clarify this question, systematic investigations with different contact geometries and/or crystal cuts have to be performed, which is out of the scope of the present work.

PbCuTe$_2$O$_6$ has a cubic chiral but non-polar crystal structure at high temperatures (space group P4$_1$32). Consequently, the onset of ferroelectricity is necessarily accompanied by the lowering of the crystal symmetry. The highest-symmetry polar subgroups of the parent cubic group are the tetragonal P4$_1$ group and the rhombohedral R3 group. Though both space groups are compatible with the presence of ferroelectricity, we expect the rhombohedral one being realized in this compound for the following reason: The cubic to rhombohedral distortion maintains the three-fold symmetry of the kagome lattice, while in case of the cubic to tetragonal distortion the kagome lattice would not be regular anymore. Thus, the former would preserve a high degree of frustration of the spin-spin interactions, while the latter would reduce it and by this likely promotes the onset of magnetic ordering, not observed experimentally. Again, future polarization measurements along different crystallographic directions should help clarifying this issue.

## Discussion

The different dielectric and lattice properties observed here for the pressed-powder material as opposed to the single crystals, can be rationalized by considering that the powder samples studied here and in the literature[18-20] were all prepared by a solid-state reaction method where the material is subject to multiple grinding processes interrupted by special heat treatments. Correspondingly, these samples constitute a more or less homogeneous collection of grains with some distribution of grain sizes, which may vary from sample to sample and the heat treatment applied. Our finding of a ferroelectric transition in the single crystalline material and the suppression of this transition in the pressed-powder samples is consistent with results on grain-size effects in ferroelectric ceramics (see, e.g., refs. 32-34 and Supplementary Note 7), yielding a critical grain size below which the transition disappears. Triggered by the results of the present work, the influence of the grain size on the 1 K phase transition in $PbCuTe_2O_6$ was systematically investigated in ref. 23. In their studies, the low-temperature specific heat on single- and polycrystalline samples was measured both after crushing the samples (thereby reducing the size of the crystallites) and after annealing them (thereby increasing the size of the crystallites and reducing dislocations and grain boundaries). According to ref. 23, the phase transition anomaly around 1 K is drastically reduced for crystallites of diameter 30 $\mu$m, and completely vanishes for diameters below 10 $\mu$m.

The drastic reduction of the anomaly in $\varepsilon'$ around 1 K for the powder sample, together with the smooth behaviour in the specific heat and thermal expansion (Fig. 1 and Fig. 2) are fully consistent with the absence of a bulk ferroelectric transition and the accompanying lattice distortion in the powder sample. At the same time, as demonstrated by various advanced magnetic measurements[19,20], these powder samples lack long-range magnetic order down to temperatures as low as 0.02 K, indicating that in the non-distorted cubic low-temperature state of PbCuTe$_2$O$_6$, the frustration of the magnetic network is strong enough to suppress long-range ordering consistent with the formation of a quantum spin liquid state. This raises the question to what extent the magnetic lattice and thus the degree of frustration is altered by the lattice distortions revealed in single crystalline material. In this context it is interesting to note that the broad maximum in the low-temperature specific heat, a feature which is considered a hallmark of strongly frustrated spin systems[13,17,18,35,36], shows a considerable shift from $T_{max} \approx 1.4$ K for the powder sample to below 1 K for the single crystalline material. It is tempting to assign this shift to alterations of the exchange coupling constants in the frustrated spin system in the single crystals due to ferroelectric ordering. This is consistent with results for the magnetization $M(T)$ derived from ac-susceptibility measurements on single crystal #5(b) (see inset of Fig. 4 and Supplementary Fig. 7 as well as Note 4), still lacking any indication for long-range order for $T \geq 0.1$ K, which reveal a mild reduction of $M(T)$ with decreasing temperatures below about 0.8 K. In contrast, for powder material a slight increase was observed for $M(T)$ on cooling below 1 K[19]. We attribute the change in the magnetic couplings to the lattice distortions - a displacement of the Pb$^{2+}$ ion out of its high-symmetry position - accompanying the ferroelectric transition, see Supplementary Fig. 1d. This displacement allows for a (*sp*) hybridization with oxygen *p* orbitals and the formation of a lone pair with asymmetric electron distribution. As these oxygen *p* orbitals are also involved in the dominant magnetic exchange paths (Supplementary Fig. 1c), we expect changes in the coupling constants $J_1$ and $J_2$ of the distorted structure.

Another even more pronounced difference in the materials' low-temperature magnetic/electronic state becomes apparent by studying the ratio $\alpha_{[110]}/C_e$ for sc #5, a quantity which is proportional to the uniaxial thermal Grüneisen parameter $\Gamma_{p[110]}$ (see ref. 37) and compare this to $\alpha/C_e$ for the pressed-powder sample pd #I, cf. inset of Fig. 5. Note, since for $T \leq 1$ K the lattice contribution to $\alpha$ is very small, its influence can be neglected in discussing the critical behaviour of the Grüneisen ratio. Figure 5 reveals a practically temperature-independent ratio for the pressed-powder sample in the temperature range investigated. This is consistent with the notion that the low-energy excitation spectrum in the powder material is governed by a single energy scale, the pressure dependence of which is probed by $\Gamma_p$. We link this energy scale to the maximum in $\alpha$ and $C_e$ at $T_{max}$ around 1.4 K which reflects magnetic correlations in the strongly frustrated hyperkagome lattice. In contrast, for the single crystal, we find a distinctly different behaviour. Whereas the ratio at higher temperatures lies close to the value for the powder material, it shows a distinctly different temperature dependence upon cooling. Of particular interest is the behaviour at lowest temperatures where the influence of the phase transition anomalies in $\alpha_{[110]}$ and $C_e$ around $T_{FE} \approx 1$ K is expected to be small, suggesting it is related to the magnetic sector. In fact, the data reveal a growing ratio $\alpha_{[110]}/C_e$ upon cooling for low temperatures. However, in light of the substantial error bars involved, the data do not allow to make a definite statement on the asymptotic behaviour, especially whether or not this ratio diverges for $T \to 0$. In general, a diverging Grüneisen ratio $\Gamma_p = 1/T \cdot (\partial T/\partial p)_S$, measuring temperature contours at constant entropy $S$, is considered a clear signature of a nearby pressure ($p$)-sensitive quantum critical point[37]. This defining property results from the accumulation of entropy close to the quantum critical point.

As an alternative approach for probing potential quantum critical behaviour associated with the system's magnetic degrees of freedom, we performed measurements of the magnetic Grüneisen parameter $\Gamma_B = 1/T \cdot (\partial T/\partial B)_S$ [37] from which the electronic magnetic Grüneisen parameter $\Gamma_{B,e}$ has been extracted by taking into account the nuclear contributions, see Methods and Supplementary Note 2 as well as Supplementary Figs. 3 and 4. In fact, as shown in the main panel of Fig. 5, the data reveal a diverging $\Gamma_{B,e}$ on cooling over about one decade in temperature down to 0.06 K with an asymptotic behaviour following a $\Gamma_{B,e} \propto T^{-m}$ dependence with $m \approx 3.9$ for $T \leq 0.25$ K. Note that an independent determination of $\Gamma_{B,e}$ is provided by using the identity $\Gamma_{B,e} = -(dM/dT)/C_e$ [37]. In fact, the calculation of $\Gamma_{B,e}$ based on the $M(T)$ data (inset of Fig. 5) and $C_e(T)$ data (Fig. 1), yields an almost identical behaviour for $\Gamma_{B,e}(T)$ within the experimental uncertainty, see Fig. 4 in the Supplementary, that is, a divergence for $T \to 0$. From these observations, we conclude that single crystalline PbCuTe$_2$O$_6$ for $T \to 0$ approaches a quantum critical point which is sensitive to magnetic fields. This is consistent with results taken on mildly increasing the magnetic field to $B = 0.2$ T (Fig. 5), and the variation of $\Gamma_{B,e}$ with field measured at a constant temperature of $T = 0.15$ K (see Supplementary Fig. 5), yielding a suppression of $\Gamma_{B,e}$ following a $\Gamma_{B,e} = -A/(B - B_c)$ dependence with $A = 0.058 \pm 0.005$ and $B_c \approx 0$. According to ref. 37 this behaviour is expected for a quantum paramagnetic state, that is, a non-critical spin-correlated state. This raises the question about the origin of the quantum critical behaviour that gives rise to the divergence in $\Gamma_{B,e}(T)$ at small fields. In light of the strong response revealed for the uniaxial thermal Grüneisen parameter along the [110] direction, we associate the quantum critical behaviour to an instability of the quantum spin liquid against uniaxial pressure $p_{[110]}$ and the associated lattice distortions. It remains to be shown by future experiments on mono-domain single crystals whether or not there are divergences in the uniaxial thermal Grüneisen parameters also along the orthogonal [1-10] and [001] directions. Remarkably, an anisotropic thermal expansion response with quantum critical behaviour only along one direction was observed for the heavy fermion metal CeRhSn[39] featuring a frustrated 2D kagome lattice.

In conclusion, comprehensive investigations on $PbCuTe_2O_6$ including thermodynamic, magnetic and dielectric probes, reveal markedly different behaviour for single crystalline material as compared to pressed-powder samples: whereas the low-temperature properties of the powder material are consistent with the recently proposed gapless quantum spin liquid state, an even more exotic behaviour is observed for the single crystals. Here, we find a ferroelectric transition at $T_{FE} \approx 1$ K, accompanied by pronounced lattice distortions, and somewhat modified magnetic signatures - still consistent with a quantum spin liquid - but with clear indications for quantum critical behaviour. These findings call for low-temperature structural investigations on single crystals as key input for determining their $T$-dependent exchange interactions. It would be interesting to extend these low-temperature structural investigations also to the pressed-powder samples, as the non-critical behaviour observed there is assigned to the lack of those lattice distortions.

In the absence of such low-$T$ structural information, we speculate that the lattice distortions in single crystalline material and the associated changes in the electronic structure are likely to alter the magnetic network in a way that drives the system close to a quantum critical point. Based on the trend revealed by comparing the maximum in the specific heat at $T_{max}$ for various frustrated antiferromagnets[13], we are inclined to assign the reduction of $T_{max}$ from 1.4 K for the powder to below 1 K for the single crystals to an increase in the degree of frustration, thereby driving the system quantum critical. It is interesting to note that quantum critical behaviour, assigned to geometrical frustration, has been observed also in other materials where a 3D frustrated spin system interacts with charge degrees of freedom such as in the Kondo lattice $Pr_2Ir_2O_7$[40] or the valence fluctuator $\beta$-$YbAlB_4$[41].

In contrast to these systems, $PbCuTe_2O_6$ features a spin system which is devoid of effects related to strong spin-orbit interactions, crystalline electric fields or Kondo-type interaction with delocalized charges. Thus, single crystalline $PbCuTe_2O_6$ offers the possibility for exploring quantum criticality resulting from a strongly frustrated hyperkagome spin system interacting with ferroelectricity – a scenario holding promise for fascinating physics. Especially, through comparison with the non-critical behaviour of the undistorted low-$T$ state realized in the pressed-powder samples of $PbCuTe_2O_6$, one may expect to find interesting magneto-electric coupling effects mediated by the lattice deformations.

## Methods

**Samples investigated.**

Single crystals of $PbCuTe_2O_6$ were grown by utilizing two different techniques: a travelling solvent floating zone (TSFZ) method for sc #1, sc #2, and sc #4, and a top-seeded solution growth (TSSG) technique for sc #5. Details of the growth conditions and the sample characterization, including polarized optical microscopy, X-ray Laue and X-ray powder diffraction, are given in ref. 23. According to these studies, the room-temperature structure of the resulting single crystals can be best refined in a cubic structure with space group P4$_1$32 and a lattice parameter $a$ = 12.4967 Å (TSSG) and 12.502 Å (TSFZ), in agreement with previous results [18]. Whereas the single crystals grown by the TSFZ technique may contain non-magnetic foreign phases with a volume fraction of 8 % at maximum, no foreign phase could be detected in these measurements for the crystals grown by using the TSSG method. The pressed-powder samples were obtained by using "polycrystalline ceramic powder" from the solid-state reaction of the precursor oxides, as described in Note 3 of the Supplementary of ref. 20. The powder was pressed into a pellet of 1 cm diameter by applying hydrostatic force up to 12 tons (~ 1.5 GPa). The starting material of the pressed-powder samples studied here, i.e., the ceramic powder of PbCuTe$_2$O$_6$, is taken from the same batch as the "polycrystalline samples" characterized and investigated via neutron scattering and synchrotron diffraction in ref. 20. According to these characterization measurements the powder is phase pure with a room-temperature structure of P4$_1$32 space group and a lattice constant $a$ = 12.4968(4) Å, see Supplementary of ref. 20, consistent with the above results for the single crystals. In addition, the influence of the preparation conditions of polycrystalline materials on the thermal properties was studied in ref. 23, and no effects of off-stoichiometry on the cation sites or oxygen deficiency could be observed.

**Thermal expansion.**

Thermal expansion experiments were performed by using a homemade capacitive dilatometer following the design discussed in ref. 42. In this technique, relative length changes $\Delta L(T)/L$ are measured with $\Delta L(T) = L(T) - L(T_0)$ and $T_0$ the starting temperature of the experiment. Measurements were performed at temperatures 0.05 K $\leq T \leq$ 2 K, using a bottom-loading $^3$He-$^4$He dilution refrigerator and for 1.5 K $\leq T \leq$ 200 K, using a $^4$He-bath cryostat. In the dilatometer cell used, the spring leafs suspending the movable part exert a force **F** to the sample along the measuring direction with $F \approx$ 0.03 N – 3 N, the actual strength of which depends on the chosen starting capacitance. This results in uniaxial pressure along the measuring direction $p_{ua} = F/A$ with $A$ the cross-sectional area of the sample, typically ranging from $p_{ua} \approx$ 0.01 MPa to 5 MPa.

**Specific heat.**

The specific heat measurements were carried out by employing both a homemade relaxation calorimeter as well as a homemade continuous-heating-type calorimeter both of which were attached to a bottom-loading $^3$He-$^4$He dilution refrigerator. In the measurements of the specific heat at low temperatures $T <$ 1.6 K, basically two different contributions have to be considered. The electronic contribution of interest here, $C_e$, arises from the magnetic moments associated with the spins of the 3d electrons of the Cu$^{2+}$ ions. The second one, $C_n$, is due to the magnetic moments of the $^{207}$Pb, $^{125}$Te, $^{63}$Cu and $^{65}$Cu nuclei present in PbCuTe$_2$O$_6$. The phonon contribution $C_{ph}$, however, is negligibly small and does not exceed 1 % of $C_e$ for $T <$ 1.6 K, see ref. 18. Below $T \approx$ 0.1 K the nuclear contribution dominates (see Note 2 and Fig. 2 of the Supplementary) and has to be determined quite accurately in order to extract $C_e$ reliably from the measured total specific heat $C$. The contributions of the various nuclei, $C_{n,i}$, are Schottky anomalies resulting from the splitting of energy levels in a magnetic field. In the case of the $^{207}$Pb and $^{125}$Te nuclei, this field is identical to the external magnetic field, as there are no electronic magnetic moments in the immediate vicinity. Thus $C_{n,i}$ can be calculated easily by using the corresponding parameters from literature (refs. 43, 44), that is, the natural abundances of the various isotopes, their nuclear spins and associated nuclear magnetic moments (see also Supplementary Table 1). For the copper nuclei, however, the local field is strongly influenced by the magnetic field caused by the magnetic moment of the 3d electrons of the $Cu^{2+}$ ion. Therefore, this local field has to be determined by a $1/T^2$ fit to the low-temperature specific heat data in $B = 0$, yielding a local electronic field of $B_e \approx 10.2$ T, see also Supplementary Note 2 for details.

**Magnetic Grüneisen parameter**.

The magnetic Grüneisen parameter, $\Gamma_B = T^{-1} \cdot (\partial T/\partial B)_S$, was measured by using a relaxation calorimeter attached to a bottom-loading $^3$He-$^4$He dilution refrigerator. For details of the applied technique, in which $\Gamma_B$ is approximated by $\Gamma_B \approx T_s^{-1} \cdot (\Delta T_s/\Delta B)_{S \approx const.}$, see ref. 38. Here $T_s$ is the sample temperature, and $\Delta T_s$ the temperature change induced in response to changes in the applied magnetic field $\Delta B$.

The electronic magnetic Grüneisen parameter $\Gamma_{B,e}$ can be extracted from the measured magnetic Grüneisen parameter $\Gamma_B$ of the total system by taking the specific heat contributions $C_i$ and magnetic Grüneisen parameters $\Gamma_{B,i}$ of all relevant subsystems into account: $\Gamma_B = \frac{1}{C} \cdot \sum_i C_i \cdot \Gamma_{B,i}$, with $C = \sum_i C_i$ the total specific heat. In the case of PbCuTe$_2$O$_6$, $\Gamma_{B,e}$ is then given by (see also Supplementary Note 2):

$$\Gamma_B = \frac{1}{C} \cdot (C_e \cdot \Gamma_{B,e} + C_{nPb} \cdot \Gamma_{B,nPb} + C_{nTe} \cdot \Gamma_{B,nTe} + C_{nCu} \cdot \Gamma_{B,nCu}) \,. \tag{1}$$

The nuclear magnetic moments of Te and Pb behave as Langevin paramagnets (for which the entropy $S = S(T/B)$) with the local magnetic field being identical to the external magnetic field, so that $\Gamma_{B,nPb} = \Gamma_{B,nTe} = 1/B$. For $\Gamma_{B,nCu}$ the situation is more complex due to the strong influence of the magnetic field caused by the magnetic moment of the 3d electrons on the local magnetic field acting on the nuclei In the absence of magnetic ordering, and for not too large magnetic fields, the following expression can be derived for the determination of $\Gamma_{B,e}$ (see Supplementary Note 2 for details):

$$\Gamma_{B,e} = \frac{1}{C_e} \cdot \left[ C \cdot \Gamma_B - (C_{nPb} + C_{nTe}) \cdot \frac{1}{B} - C_{nCu} \cdot \frac{B}{3B_e^2} \right] \,. \tag{2}$$

**Dielectric measurements.**

For measurements of the dielectric constant, a plate capacitor arrangement was realized by attaching two electrodes (silver paste) to opposite parallel surfaces of the samples. The dielectric constant was derived from the capacitance, read out by using an LCR meter (Agilent E4980), and the geometrical dimensions of the plate capacitor. This procedure implies an uncertainty in the experimental data of about ± 5 %. Measurements were performed by using a top-loading $^3$He-$^4$He dilution refrigerator.

**Polarization.**

The polarization measurements were performed using the same capacitor-contact geometry as described for the dielectric measurements. To determine the ferroelectric polarization, standard pyrocurrent experiments were performed by prepoling the sample upon cooling down to 0.9 K with different applied electrical fields and subsequent monitoring of the pyrocurrent under heating beyond the transition temperature. The pyrocurrent was recorded using a Keysight B2987A electrometer and applying a heating rate of 60 mK/min. For examples of pyrocurrent measurements see Note 8 and Fig. 10 of the Supplementary. The polarization was deduced by integration of the pyrocurrent over time. To check for the switchability of the polarization, additional experiments were performed by (i) measuring the pyrocurrent upon cooling the sample with applied electrical field, (ii) subjecting the sample to a field of opposite direction for 1 min after reaching a base temperature of 0.9 K, and (iii) monitoring the pyrocurrent upon subsequent heating in zero field. Typical fields applied were of the order of several kV/cm.

**Ac-susceptibility.**

For measurements of the ac-susceptibility, a homemade susceptometer adapted to a top-loading $^3$He-$^4$He dilution refrigerator was employed. The ac-susceptometer was calibrated via magnetization measurements up to 5 T by comparing the results with data obtained by using a SQUID magnetometer (Quantum Design MPMS).

## Data availability

All the raw and derived data that support the findings of this study are available from the authors upon reasonable request.

## Acknowledgements

Work done at Goethe University Frankfurt was supported by the German Science Foundation (DFG) through the SFB/TR 288 (ID 422213477). Work done at the University of Augsburg was supported by the DFG via the SFB/TR 80 (ID 107745057). B.L., A.R.N.H. and A.T.M.N. I. acknowledge the support of DFG through project B06 of SFB 1143 (ID 247310070). We acknowledge fruitful discussions with Christian Hess, Xiao-Chen Hong, Elena Gati and Harald O. Jeschke.

## Author contributions

B.L., S.C., B.W. and M.L. planned the project. Measurements and analysis of specific heat and magnetic Grüneisen parameter were performed by P.E. and U.T. Measurements and analysis of thermal expansion were performed by C.T., Y.S., S. H., B.W. and M.L. Measurements and analysis of the magnetic susceptibility were performed by B.W. and P.E. Measurements and analysis of the dielectric constant were performed by A.A. and U.T. Measurements and analysis of ultrasonic experiments were performed by J.Z. and B.W. Measurements of electric polarization were performed by M.W., B.W. and A.A. The data were discussed and analysed by M.W., P.L. and I.K. The samples were synthesized and characterized by A.R.N.H.,

# Figures

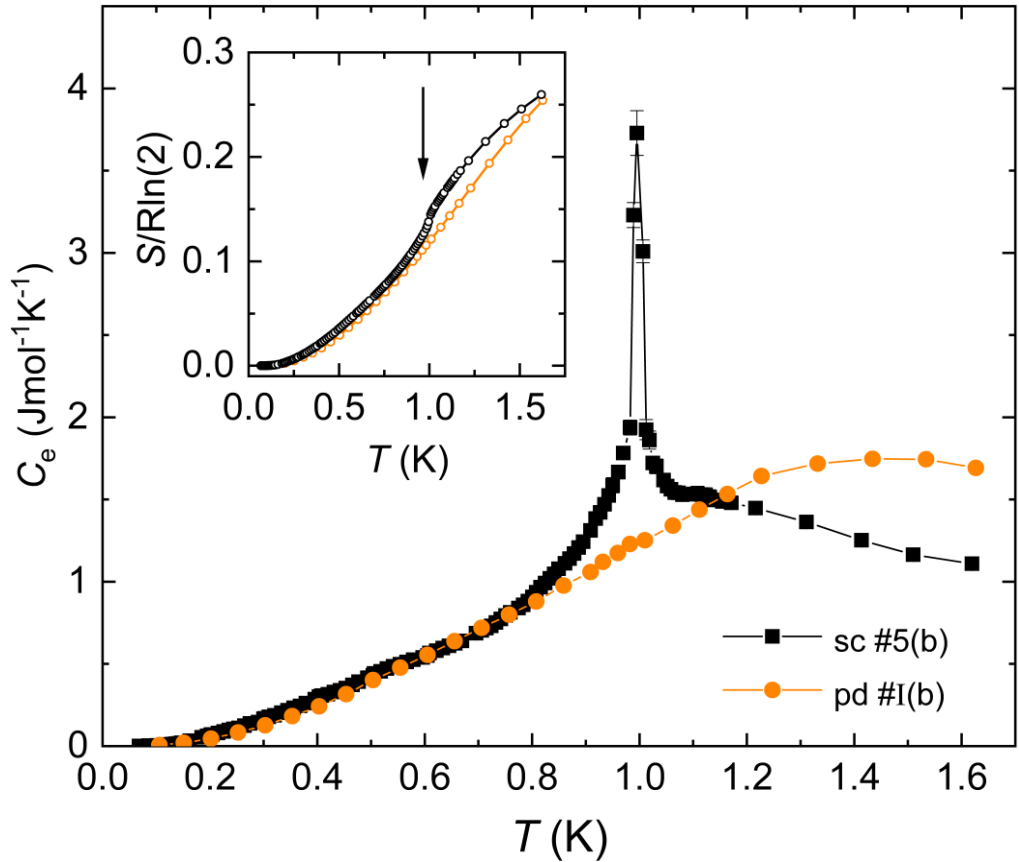

**Figure 1 Temperature dependence of the specific heat of PbCuTe$_2$O$_6$.** Electronic specific heat of single crystal sc #5(b) (black squares) and a pressed-powder sample pd #I(b) (orange spheres). Thin solid lines serve as guides to the eyes. Very similar results to those on sc #5(b) were obtained on a second single crystal #1(b) studied in this experiment, which was grown using a different technique, see Methods and Supplementary Fig. 2. For the nuclear contributions subtracted, see Methods and Supplementary Fig. 2 and Note 2. Error bars represent the standard deviation due to statistical treatment of raw data. The inset shows the entropy in units of $R\ln2$ for both samples using the same colour code. The arrow marks the position of the phase transition anomaly in $C_e(T)$. The entropy release associated with the phase transition over a narrow temperature window of about 0.2 K amounts to a few percent of $R\ln2$.

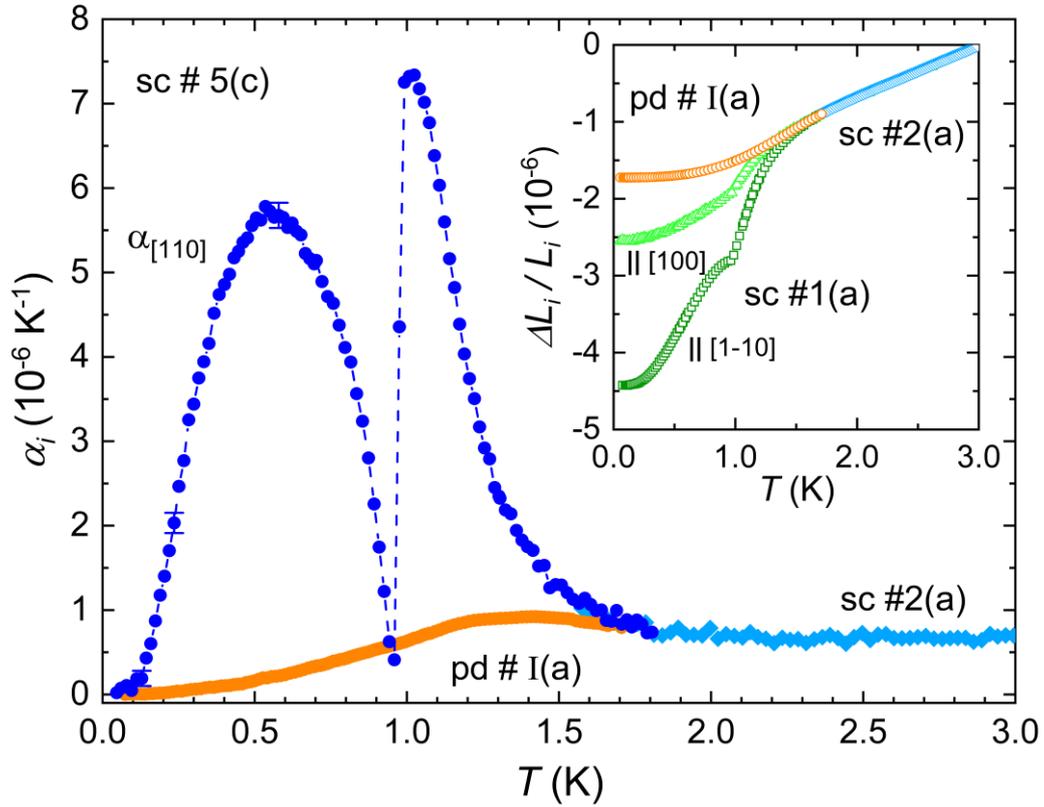

**Figure 2 Temperature dependence of the coefficient of thermal expansion of PbCuTe$_2$O$_6$.** Thermal expansion coefficient $\alpha$ for single crystal sc #5(c) (dark blue spheres) and sc #2 (light blue diamonds for $T \geq 1.5$ K), both measured along the [110] direction, together with data for the pressed-powder sample pd #I(a) (orange spheres). Broken line serves as a guide to the eyes. Representative error bars resulting from the statistical treatment of raw data and systematical uncertainties related to the determination of the sample length. Error bars of measurement for pd sample are within symbol size. The inset shows relative length changes $\Delta L_i(T)/L_i$ for sc #1(a), measured along [100] (light green triangles) and [1-10] (dark green squares), sc #2 (light blue diamonds) for $T \geq 1.5$ K, and for the powder sample pd # I(a) (orange spheres). The pronounced anomaly in $\alpha_{[110]}$ for $T \leq 1.5$ K is consistent with a strong decrease observed in the longitudinal elastic constant $c_{L[110]}$ upon cooling, see Fig. 8 and Note 5 in Supplementary.

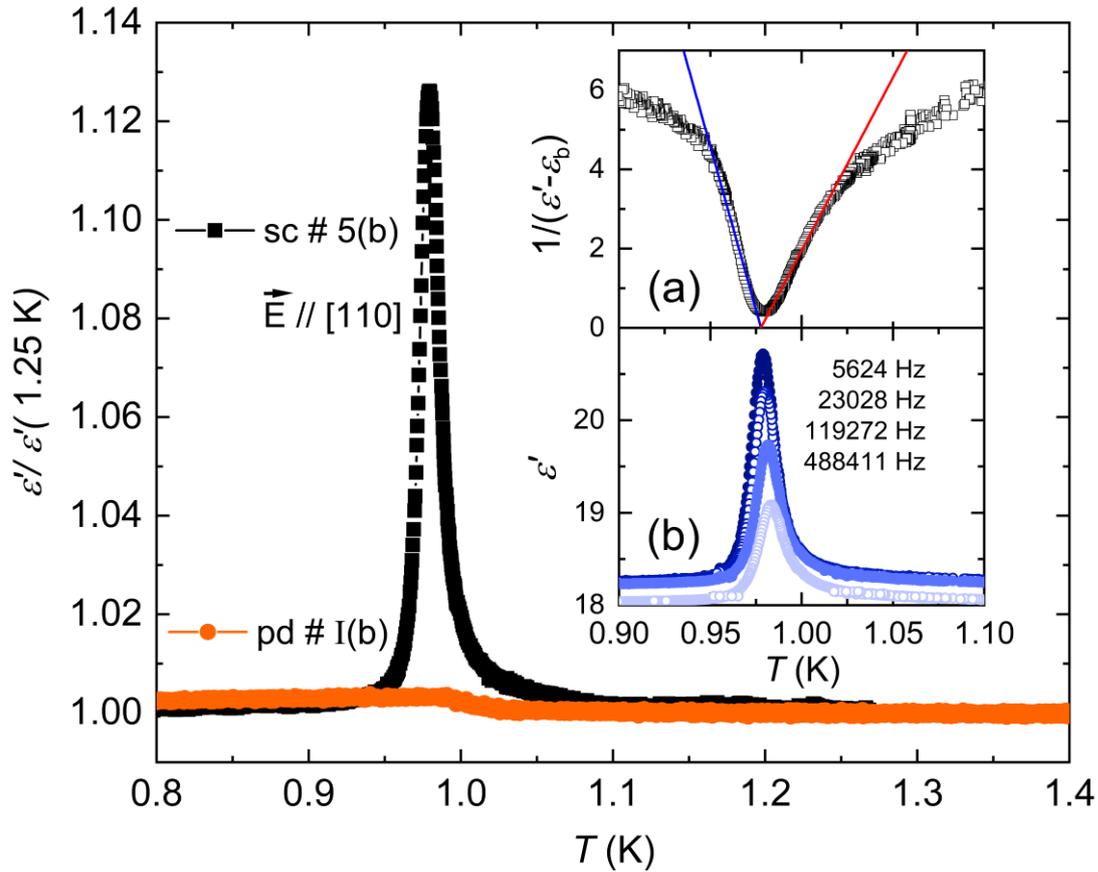

**Figure 3 Temperature dependence of the dielectric constant of PbCuTe$_2$O$_6$.**

Dielectric constant of single crystal sc #5(b) (black squares), with electric field **E** applied along the [110] direction, and a pressed-powder sample pd #I(b) (orange spheres). Similar measurements on the single crystals #1(a) and #4 reveal essentially the same results albeit with a somewhat reduced peak height, see Supplementary Fig. 9 and Note 6. The inset gives details of the data for sc #5(b) in a narrow temperature range around 1 K. Inset (a), showing $1/(\varepsilon' - \varepsilon'_b)$ vs. $T$, reflects a Curie-Weiss-like behaviour on approaching the maximum in $\varepsilon'$ from both sides. The slopes of the linear regimes differ by a factor of about -1.9. This factor is very close to -2, expected for a mean-field second-order ferroelectric transition[25], but strongly departs from -8 seen at first-order ferroelectric transitions[26,27]. The lower inset (b) shows $\varepsilon'$ vs. $T$ for varying frequencies as indicated in the figure. The small shift of the background dielectric constant $\varepsilon'_b$, visible for the data at highest frequency, indicates some contribution from extrinsic factors.

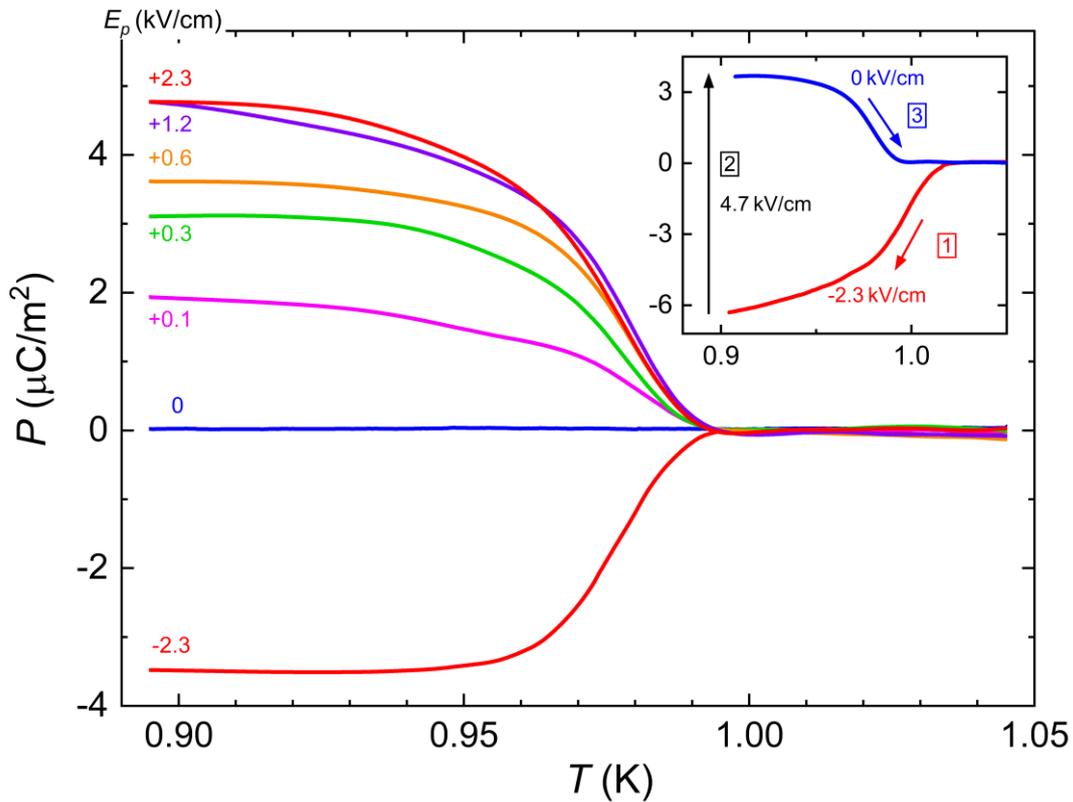

**Figure 4 Electric polarization of single crystalline PbCuTe$_2$O$_6$.** Temperature dependence of the electric polarization of single crystalline PbCuTe$_2$O$_6$ (sc #1(a)) as deduced from the pyrocurrent monitored upon heating. The measurements were performed for different poling fields (-2.3 to +2.3 kV/cm), applied to the sample along the [110] direction during the preceding cooling run. For original pyrocurrent data taken at different heating rates and poling fields, see Supplementary Fig. 10a and Fig. 10b. The results in the inset demonstrate the switching of the polarization achieved by first cooling the sample within negative field (1), applying a positive field for 1 min at 0.9 K (2), and finally reheating it without field (3).

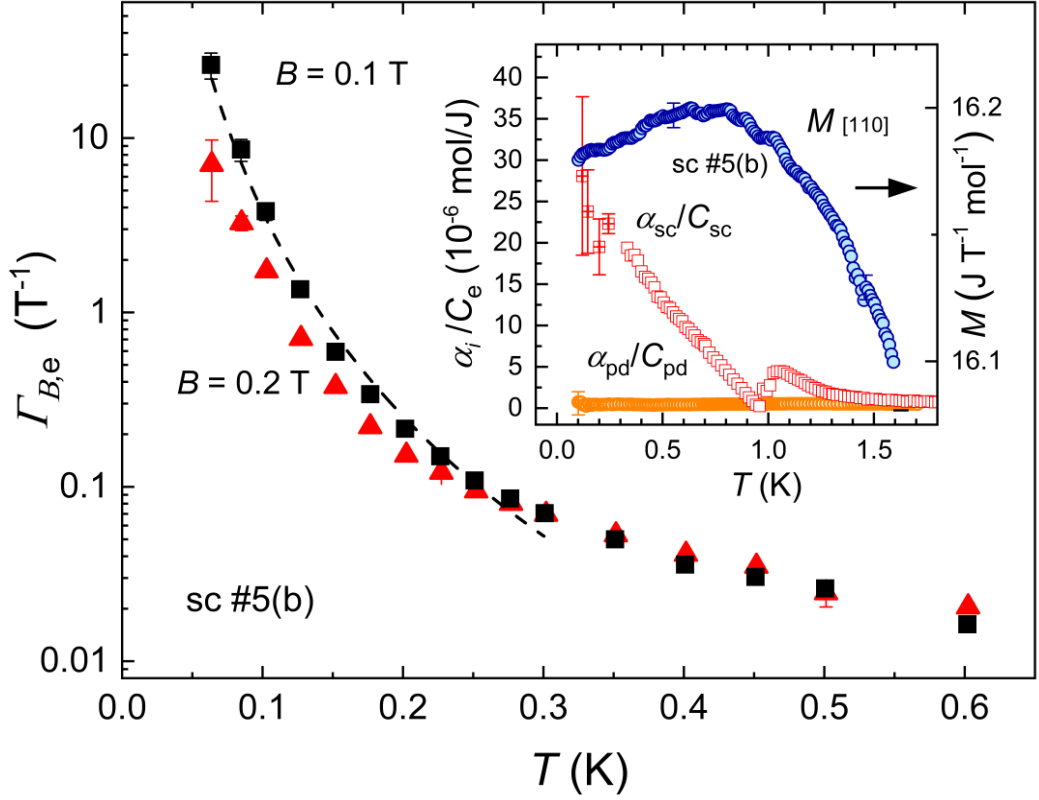

**Figure 5 Temperature dependence of the magnetic Grüneisen parameter.**

Electronic magnetic Grüneisen parameter $\Gamma_{B,e}$ as a function of temperature for PbCuTe$_2$O$_6$ single crystal #5(b) in a semi-logarithmic representation. Raw data were taken by applying small field ramps of $\Delta B = \pm\, 0.03$ T on top of a dc field of $B = 0.1$ T (black squares) and 0.2 T (red triangles) applied along [110], see Methods and Supplementary Note 2 as well as ref. 38 for details on the technique applied. A minimum dc field of 0.1 T was used to avoid issues associated with remnant fields of the magnet. Broken line shows a fit $T^{-m}$ with $m = 3.9$ to the data for $T \leq 0.25$ K. Error bars result from averaging over field-up and field-down sweep. The inset (left ordinate) shows the ratio $\alpha_{sc}/C_{sc} = \alpha_{[110]}/C_{sc}$ for single crystal #5 (open red squares) and $\alpha_{pd}/C_{pd}$ for the pressed-powder sample pd #I (open orange spheres) both measured at $B = 0$. Errors are calculated from the errors of $\alpha$ and $C$. The inset also shows results for the magnetization $M(T)$ (right ordinate) per mole in a field of $B = 0.1$

T along the [110] direction of sc #5(b) obtained by ac-susceptibility measurements. Representative error bar, denoting standard deviation due to statistical treatment of raw data.

Supplementary Information

"Spin liquid and ferroelectricity close to a quantum critical point in PbCuTe$_2$O$_6$"

Christian Thurn[1,*], Paul Eibisch[1], Arif Ata[1], Maximilian Winkler[2], Peter Lunkenheimer[2], István Kézsmárki[2], Ulrich Tutsch[1], Yohei Saito[1], Steffi Hartmann[1], Jan Zimmermann[1], Abanoub R. N. Hanna[3,4], A. T. M. Nazmul Islam[4], Shravani Chillal[4], Bella Lake[3,4], Bernd Wolf[1], Michael Lang[1]

[1]Physikalisches Institut, J.W. Goethe-Universität Frankfurt(M), 60438 Frankfurt, Max-von-Laue-Str.1, Germany

[2]Experimentalphysik V, Zentrum für Elektronische Korrelationen und Magnetismus, Universität Augsburg, Universitätsstr. 1, 86159 Augsburg, Germany

[3]Institut für Festkörperforschung, Technische Universität Berlin, Hardenbergstr. 36, 10623 Berlin, Germany

[4]Helmholtz-Zentrum Berlin für Materialien und Energie, Hahn-Meitner Platz 1, 14109 Berlin, Germany

*e-mail: Thurn@Physik.uni-frankfurt.de

# Supplementary Notes

## 1. Crystal structure, structural motives and magnetic as well as dielectric units of $PbCuTe_2O_6$

$PbCuTe_2O_6$ crystallizes in a non-centrosymmetric cubic structure with space group $P4_132$ (No. 213), see Fig. 1. According to density functional theory calculations [1] based on room-temperature structural data, the magnetic lattice can be described by isolated equilateral $S = 1/2$ triangles with nearest-neighbour interaction $J_1 = 1.13$ meV, which are coupled via the second-nearest neighbour interaction $J_2 = 1.07$ meV into a hyperkagome lattice (Fig. 1b). The weaker third- and fourth-nearest neighbour interactions $J_3 = 0.59$ meV and $J_4 = 0.12$ meV couple the spins into chains. Another unique feature of $PbCuTe_2O_6$, which has been largely ignored until now, relates to its dielectric degrees of freedom [2]. The material contains polar building blocks originating from the free electron pairs (lone pairs) of the $Te^{4+}$ ions in the oxotellurate tetrahedrons and the asymmetrically coordinated $Pb^{2+}$ ions. These characteristics together with the non-centrosymmetric structure imply the possibility of ferroelectric order interacting with the strongly frustrated quantum spin system.

## 2. Specific heat and magnetic Grüneisen parameter

At low temperatures ($T < 1.6$ K) there are basically two different contributions to the specific heat, $C$, of $PbCuTe_2O_6$. The first one, $C_e$, arises from the magnetic moments associated with the spins $\vec{S}$ of the 3d electrons of the $Cu^{2+}$ ions, and the second one, $C_n$, is due to the magnetic moments of the $^{207}Pb$, $^{125}Te$, $^{63}Cu$ and $^{65}Cu$ nuclei. The phonon contribution $C_{ph}$ [3], however, is negligibly small and does not exceed 1 % of $C_e$ for $T < 1.6$ K. Below $T \approx 0.1$ K the nuclear contribution dominates and has to be determined quite accurately in order to extract $C_e$ reliably from the measured total specific heat $C$.

The contributions $C_{n,i}$ of the above listed nuclei to the specific heat are Schottky anomalies with equidistant energy levels produced by the different orientations of the nuclear magnetic moments $\vec{\mu}_i$ in the local magnetic fields $\vec{B}_i^*$:

$$C_{n,i} = z_i \cdot \frac{R}{2} \cdot \left(\frac{\Delta_i}{k_B T}\right)^2 \cdot \left(\frac{1}{\cosh\left(\frac{\Delta_i}{k_B T}\right)-1} - \frac{a_i^2}{\cosh\left(\frac{a_i \Delta_i}{k_B T}\right)-1}\right) . \quad (1)$$

Here $z_i$ denotes the number of nuclei of type $i$ in one formula unit, $a_i = 2 \cdot I_i + 1$ (with $I_i$ the nuclear spin quantum number) is the number of energy levels of such a nucleus in the local magnetic field $\vec{B}_i^*$ and $\Delta_i = \mu_i \cdot B_i^*/I_i$ (where $B_i^* = |\vec{B}_i^*|$, $\mu_i = |\vec{\mu}_i|$) describes the spacing between neighbouring energy levels of a nucleus of type $i$. In the case of the $^{207}$Pb and $^{125}$Te nuclei $\vec{B}_i^*$ is identical to the external magnetic field $\vec{B}$, as there are no electronic magnetic moments in the vicinity which would produce a (significant) disturbance of the magnetic field. Thus $C_{n,i}$ can be calculated easily by using the corresponding parameters from the table 1 below [4,5]. For the copper nuclei, however, $\vec{B}_i^*$ is strongly influenced by the magnetic field $\vec{B}_e$ from the magnetic moment of the 3d electrons of the $Cu^{2+}$ ion. Therefore, $B_e = |\vec{B}_e|$ has to be determined by asymptotically fitting the low temperature specific heat $C$ in $|\vec{B}| = B = 0$ with the formula (1) using the corresponding parameters of the copper nuclei and leaving $B^*_{Cu63} = B^*_{Cu65} = B_e$ as free parameter (the other nuclei do not contribute in this case, as they do not experience a local magnetic field. The fit is shown in the inset of Fig. 2 and yields the value $B_e$ = 10.2 T.

The magnetic Grüneisen parameter $\Gamma_{B,e}$ of the electron spins can be extracted from the measured magnetic Grüneisen parameter $\Gamma_B = T^{-1} \cdot (\partial T/\partial B)_S$ of the total system by taking the specific heat contributions $C_i$ and magnetic Grüneisen parameters $\Gamma_{B,i}$ of all relevant subsystems into account. In general, the following relation holds:

$$\Gamma_B = \frac{1}{C} \cdot \sum_i C_i \cdot \Gamma_{B,i} \quad (2)$$

with $C = \sum_i C_i$ the total specific heat. In the case of PbCuTe$_2$O$_6$, $\Gamma_{B,e}$ is then given by

$$\Gamma_B = \frac{1}{C} \cdot (C_e \cdot \Gamma_{B,e} + C_{nPb} \cdot \Gamma_{B,nPb} + C_{nTe} \cdot \Gamma_{B,nTe} + C_{nCu} \cdot \Gamma_{B,nCu}). \tag{3}$$

The nuclear magnetic moments of tellurium and lead behave as Langevin paramagnets (for which the entropy $S = S(T/B)$) with the local magnetic field $\vec{B}_i^*$ being identical to the external magnetic field $\vec{B}$, so that

$$\Gamma_{B,nPb} = \Gamma_{B,nTe} = \frac{1}{B}. \tag{4}$$

For $\Gamma_{B,nCu}$ the situation is less simple due to the strong influence of $\vec{B}_e$ on $\vec{B}_i^*$. As the magnetic moments of the copper nuclei practically do not interact with one another, their $T/B_i^*$ ratio is constant in an adiabatic process:

$$\frac{T}{B_i^*} = const. \tag{5}$$

The local field is given by

$$B_i^* = \sqrt{B_e^2 + B^2 + 2B_e B \cos\vartheta}, \tag{6}$$

with $\vartheta$ being the angle between $\vec{B}$ and $\vec{B}_e$. Thus the following expression is obtained

$$\Gamma_{B,Cu} = \frac{1}{T} \cdot \left(\frac{\partial T}{\partial B}\right)_S = \frac{B + B_e \cos\vartheta}{B_e^2 + B^2 + 2B_e B \cos\vartheta}. \tag{7}$$

This can be simplified for the case $B \ll B_e$ by a series expansion and taking into account only the terms which are linear in $B/B_e$:

$$\Gamma_{B,nCu} = \frac{1}{B_e} \cdot \left[\cos\vartheta + (1 - 2\cos^2\vartheta) \cdot \frac{B}{B_e}\right]. \tag{8}$$

As the copper spins show no indications of ordering in a small external magnetic field $\vec{B}$, they can be assumed to be oriented completely randomly relatively to $\vec{B}$ and thus the same holds for $\vec{B}_e$ too. Therefore, (8) has to be averaged evenly over the whole solid angle:

$$\overline{\Gamma}_{B,\text{nCu}} = \frac{1}{4\pi} \cdot \int \Gamma_{B,\text{nCu}} d\Omega = \frac{B}{3B_e^2} \quad . \tag{9}$$

Substituting (4, 9) into (3) and solving the equation for $\Gamma_{B,e}$ yields the final formula

$$\Gamma_{B,e} = \frac{1}{C_e} \cdot \left[ C \cdot \Gamma_B - (C_{\text{nPb}} + C_{\text{nTe}}) \cdot \frac{1}{B} - C_{\text{nCu}} \cdot \frac{B}{3B_e^2} \right] \quad . \tag{10}$$

Figure 2 shows the specific heat contribution, $C_e$, associated with the $Cu^{2+}$ electron spins of $PbCuTe_2O_6$ for two single crystals sc #1(b) and #5(b). The specific heat for both crystals is almost identical around the phase transition anomaly and for temperatures below. Some differences are visible for temperatures above the phase transition where $C(T)$ for #1(b) is slightly higher than $C(T)$ for #5(b).

(1) The first technique applied [6] follows the definition $\Gamma_B = T^{-1}(\partial T/\partial B)_S$, by measuring temperature changes $\Delta T^*$ of the sample, which is in weak thermal contact to a bath at temperature $T_b$, in response to changes of the magnetic field $\Delta B$, i.e., $\Gamma_B \approx T^{-1} \cdot (\Delta T^*/\Delta B)_{S\approx\text{const.}}$. A typical measuring cycle is shown in Fig. 3. Since this technique measures the total Grüneisen parameter $\Gamma_B$, which is the sum of various contributions (see eq. (3)), the determination of the electronic part $\Gamma_{B,e}$ requires a careful consideration of the nuclear contributions according to eq. (10).

2) As an alternative approach, we took advantage of the identity $\Gamma_{B,e} = -C_e^{-1} \cdot (\partial M/\partial T)$ by using the electronic specific heat data, $C_e$, of sc #5(b) shown in Fig. 1 of the main text, and the magnetization, $M(T)$, of sc #5(b) displayed in the inset of Fig. 5 of the main text. The so-derived $\Gamma_{B,e}$ is also shown in Fig. 4 (blue spheres). As the figure demonstrates, we find a good agreement between the results obtained by following these different approaches. In particular, both approaches reveal a divergence of $\Gamma_{B,e}(T)$ for $T \to 0$.

## 3. Thermal expansion

Figure 6 shows the coefficient of thermal expansion of PbCuTe$_2$O$_6$ single crystals sc #1(a) along the [1-10] and #sc 5(c) along the [110] direction for $T \leq 2$ K.

In the dilatometer cell used (constructed following the design discussed in [7]), the spring leafs, suspending the movable part, exert a force **F** to the sample along the measuring direction, with $F \approx 0.03$ N – 3 N, the actual strength of which depends on the chosen starting capacitance $C_s$. This results in a uniaxial pressure along the measuring direction $p_{ua} = F/A$ with $A$ the cross-sectional area of the sample, typically ranging from $p_{ua} \approx 0.01$ MPa – 5 MPa.

The data in Fig. 6 represent two data sets taken along an in-plane diagonal: the data for sc #1(a) (green squares) were taken with a uniaxial pressure of moderate strength $p_{ua} = (2.3 \pm 0.5)$ MPa, whereas for sc #5(c) (blue squares) the uniaxial pressure was relatively high with $p_{ua} = (6.5 \pm 1.3)$ MPa. The pressure values refer to values at low temperatures around 4 K. The data sets show practically identical behaviour at the high-temperature end, i.e., from around 2 K down to about 1.4 K. For lower temperatures, both data sets show anomalous behaviour, which is qualitatively similar, albeit with different amplitudes. We assign these differences to the influence of the uniaxial pressure on the domain structure associated with the formation of ferroelectric order around 1 K and the accompanying lattice distortions to a non-cubic low-temperature state. The data for sc #5(c) demonstrate that for a preferential domain orientation – or even a mono-domain configuration – realized here, the anomaly in the coefficient of thermal expansion accompanying the ferroelectric order becomes more pronounced.

## 4. Magnetic susceptibility

The main panel of figure 7 exhibits the ac-susceptibility of crystal #5(b) as a function of temperature for 0.1 K ≤ $T$ ≤ 1.6 K in a magnetic field of 0.1 T. The data were taken by using a home-made ac-susceptometer adapted to a top-loading dilution refrigerator. The ac-susceptometer was calibrated via magnetization measurements up to 5 T by comparing the results with data obtained by using a SQUID magnetometer (Quantum Design MPMS). For the $\chi_{ac}$ measurements the external field was aligned along the [100] (red filled circles, right scale) and [110] (blue filled circles, left scale) directions. Unlike specific heat, thermal expansion and dielectric experiments, all of which showing pronounced phase transition anomalies around 1 K, no indications for a phase transition are visible in the magnetic response. The g-factor anisotropy for the two field orientations amounts to approximately 5 %, an anisotropy which is typical for $Cu^{2+}$ in a square-planar environment [8]. In the inset the results from the ac-susceptibility measurements for $T$ ≤ 1.6 K are shown together with the static susceptibility data of sc # 5(b) (red and blue squares) taken by using a commercial Quantum Design SQUID magnetometer in the temperature range 2.0 K ≤ $T$ ≤ 20.0 K for the same field orientations. The data reveal a small anomaly around 6 K of unknown origin. A more pronounced anomaly is also visible in the same temperature range for sc #2 (not shown). In this sample sc #2 there is also a small but distinct difference visible between field-cooled and zero-field cooled data. Similar anomalies in this temperature range were also observed in other frustrated triangular-lattice systems and were ascribed to a spin freezing induced by quenched disorder [9].

## 5. Elastic constant

Figure 8 exhibits the thermal expansion $\alpha_{[110]}$ (orange circles, sc #2) and the longitudinal elastic constant $c_{L[110]}$ (blue squares, sc #5(a)) both as a function of temperature for 1.3 K ≤ $T$ ≤ 20.0 K. The elastic constant was also measured along the [110] direction. The thermal expansion of sample sc #2 continuously decreases and exhibits a small hump-like anomaly around 7 K. With further decreasing temperature there is a significant increase of $\alpha_{[110]}$ below 2.2 K which is the precursor of the ferroelectric transition at the $T_{FE} \approx 1$ K, cf. Fig. 2 in the main text. For the longitudinal elastic constant along the [110] direction, we observe a continuous increase of $c_{L[110]}$ upon cooling from room temperature down to approximately 10 K where it adopts a broad maximum. Below 10 K $c_L$ decreases moderately strongly down to 1.3 K, the lowest temperature of our experiment. Below approximately 2 K the softening becomes stronger which we interpret - in analogy to the thermal expansion - as a precursor of the ferroelectric transition.

## 6. Dielectric constant

The dielectric constant $\varepsilon'$ was measured on 3 single crystals of PbCuTe$_2$O$_6$ in the temperature range 0.25 K $\leq T \leq$ 1.3 K, see Figure 9. In these measurements the electrical field **E** was oriented parallel to the [110] direction for sc #5(b) and sc #1(a), and parallel to the [001] direction for sc #4. For the determination of the dielectric constant $\varepsilon'$, a plate capacitor geometry was realized by attaching two electrodes (silver paste) to opposite parallel surfaces of the samples. The dielectric constant was then derived from the capacitance, read out by using an LCR meter (Agilent E4980), and the geometrical dimensions of the plate capacitor. This procedure implies an uncertainty of about ± 5 %. For all single crystals investigated, we observe a peak in $\varepsilon'$ with small variations in the peak position at (0.97 ± 0.05) K. There is, however, some variation in the size of the peaks, even for the same orientation of the electrical field, indicating some sample-to-sample dependence. The relatively narrow peak sits on top of a temperature-independent background contribution $\varepsilon_b'$. We find similarly high values around $\varepsilon_b' \approx 18$ for sc #5(b) and sc #1(a) for **E** parallel [110] and an even higher value of $\varepsilon_b' \approx 23$ for sc #4 where **E** was parallel to [001]. In contrast, for the pressed-powder sample pd #I(b), we find a reduced $\varepsilon_b' \approx 12$ and only a tiny anomaly around 1 K. From the size of the anomaly and the fact that thermal expansion measurements failed to detect any lattice distortion associated with it (see Fig. 2 main text), we claim that there is no bulk ferroelectric transition in the investigated pressed-powder samples.

## 7. Grain-size effect of ferroelectric order

The observation made here of a ferroelectric transition in PbCuTe$_2$O$_6$ single crystals and the suppression of this transition in pressed-powder samples is consistent with results on grain-size effects in ferroelectric ceramics [10-12], yielding a critical grain size below which the transition disappears. Whereas for an isolated grain the instability of the ferroelectric phase is mainly due to the surface effect, the situation becomes more complex for ceramics where beside intrinsic grain-size effects other factors, which may change with the size of the system, can be of relevance as well, see ref. [12] and references cited therein.

## 8. Pyrocurrent

Figure 10a provides examples of the pyrocurrent results used to deduce the temperature-dependent electric polarization curves shown in Fig. 4 as described in the methods section of the main text. Typical results for two different positive prepoling fields with a standard heating rate of 60 mK/min, used for all measurements in Fig. 4, are shown. The peak revealed at the transition temperature reflects the successive reorientation of the poled electrical dipoles into disordered equilibrium positions upon heating. As expected, the pyrocurrent is enhanced for higher poling field, corresponding to stronger polarization.

Figure 10b shows the pyrocurrent for two different heating rates and identical, negative prepoling fields. As expected, the current is enhanced for higher rate. For conducting materials, thermally stimulated discharge currents can also lead to pyrocurrent peaks.

The fact that the peak temperature in Fig. 10b is unaffected by the heating rate excludes this non-intrinsic effect, which should lead to a significant shift of the peak to higher temperatures for higher heating rates [13,14].

**Supplementary Figures**

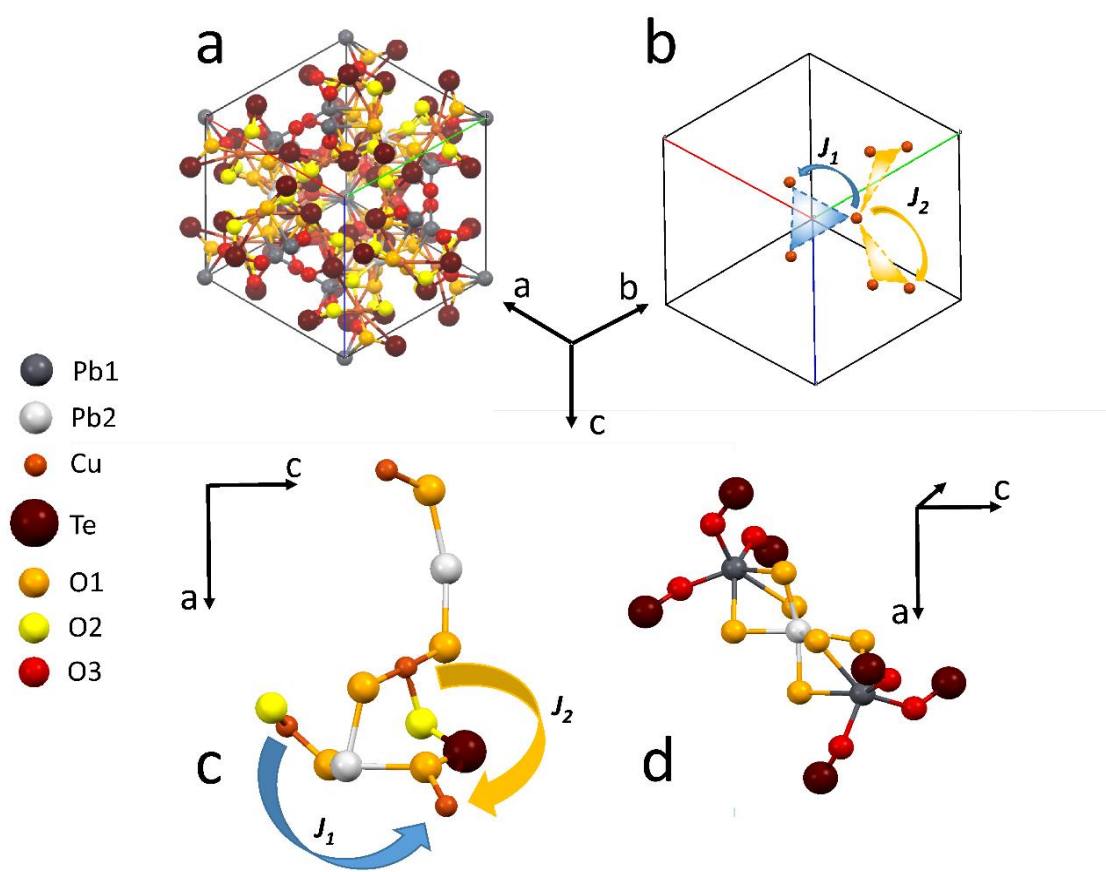

**Figure 1**. Crystal structure, structural motives and functional units of PbCuTe$_2$O$_6$. For the atoms in all subfigures a colour code is used as indicated in the figure: Pb1 dark grey, Pb2 light grey, Cu dark orange, Te brown, O1 dark yellow, O2 yellow and O3 red.

a: View along the [111]-direction of the cubic high-temperature structure with space group P4$_1$32. The figure shows one unit cell containing 12 crystallographically equivalent magnetic Cu$^{2+}$ ions with $S = 1/2$ forming a three-dimensional (3D) geometrically frustrated hyperkagome lattice.

b: Reduced magnetic structure containing the key motifs. These are equilateral triangles formed by the Cu$^{2+}$ ($S = 1/2$) ions representing the two dominant 1$^{st}$ (blue) and 2$^{nd}$ (yellow) nearest-neighbour interactions with corresponding magnetic coupling constants $J_1$ and $J_2$ [1]. Whereas the blue triangles form isolated magnetic units, the yellow ones build a 3D hyperkagome network (not shown here; see, e.g., Fig. 1 of ref. 1) of corner-sharing triangles.

c: Magnetic exchange paths corresponding to $J_1$ and $J_2$ for the isolated triangles (blue arrow) and the 3D network of corner-sharing triangles (yellow arrow), respectively. Whereas $J_1$ (blue arrow) is mediated via the Cu-O1-Pb2-O1-Cu orbital overlap, $J_2$ (yellow arrow) results from the Cu-O2-Te-O1-Cu orbital overlap.

d: The dielectric unit consists of chiral tellurate complexes and the Pb1$^{2+}$ and Pb2$^{2+}$ ions possessing stereochemically active lone pairs of their 6s electrons. These dielectric units form chains along the [111]-direction, a three-fold rotation axis of the cubic high-temperature structure. The Pb1 ions are in a distorted oxygen coordination with three short bonds to the O3 ions and three long bonds to O1 ions. These long bonds provide a void for the lone pair electrons of Pb1. In contrast, there is no lone pair associated with Pb2 in its symmetrically coordinated environment, forming bonds of similar length to six O1 ions. Therefore, in the non-distorted high-temperature structure the dipole moments originating from the tellurates and Pb1 within each unit compensate each other. Thus a distortion along the three-fold rotation axis, i.e., a displacement of the Pb2 out of its high-symmetry position, is necessary to allow for a (sp) hybridization [15] resulting in the formation of a lone pair with asymmetric electron distribution prerequisite to the formation of ferroelectric order. This process can be described as a second-order Jahn-Teller instability [15]. As the O1 $p$ orbitals are involved in the dominant magnetic exchange paths, we expect that the distorted structure (with the modified electron distribution) will also result in a change in the magnetic coupling constants $J_1$ and $J_2$.

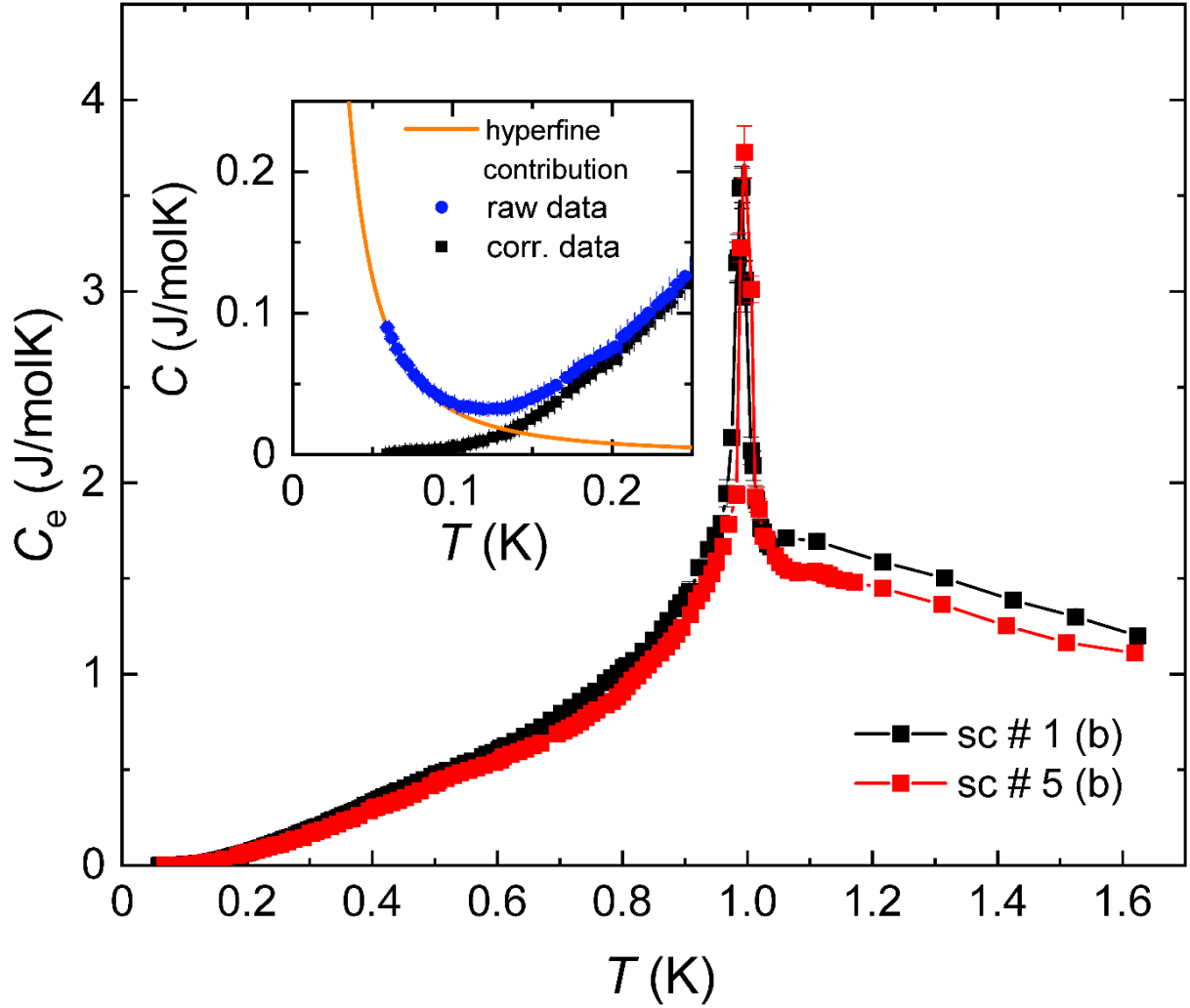

**Figure 2**. Electronic contribution to the specific heat, $C_e$, related to the $Cu^{2+}$ electron spins of $PbCuTe_2O_6$ for single crystals #1(b) and #5(b). The inset shows the total specific heat, $C$, of crystal #1(b) (blue circles) together with the asymptotical low-temperature fit to these data representing the nuclear contribution, $C_n$, associated with the copper isotopes $^{65}Cu$ and $^{63}Cu$ (orange line). As in the main graph the black squares denote the specific heat $C_e$ of the $Cu^{2+}$ electron spins after subtracting the nuclear contribution $C_n$.

The magnetic Grüneisen parameter of interest $\Gamma_{B,e} = -C_e^{-1}(\partial S_e/\partial B)_T$, which probes the variation of electronic entropy, $S_e$, with respect to changes of the magnetic field under isothermal conditions, was measured in two different ways.

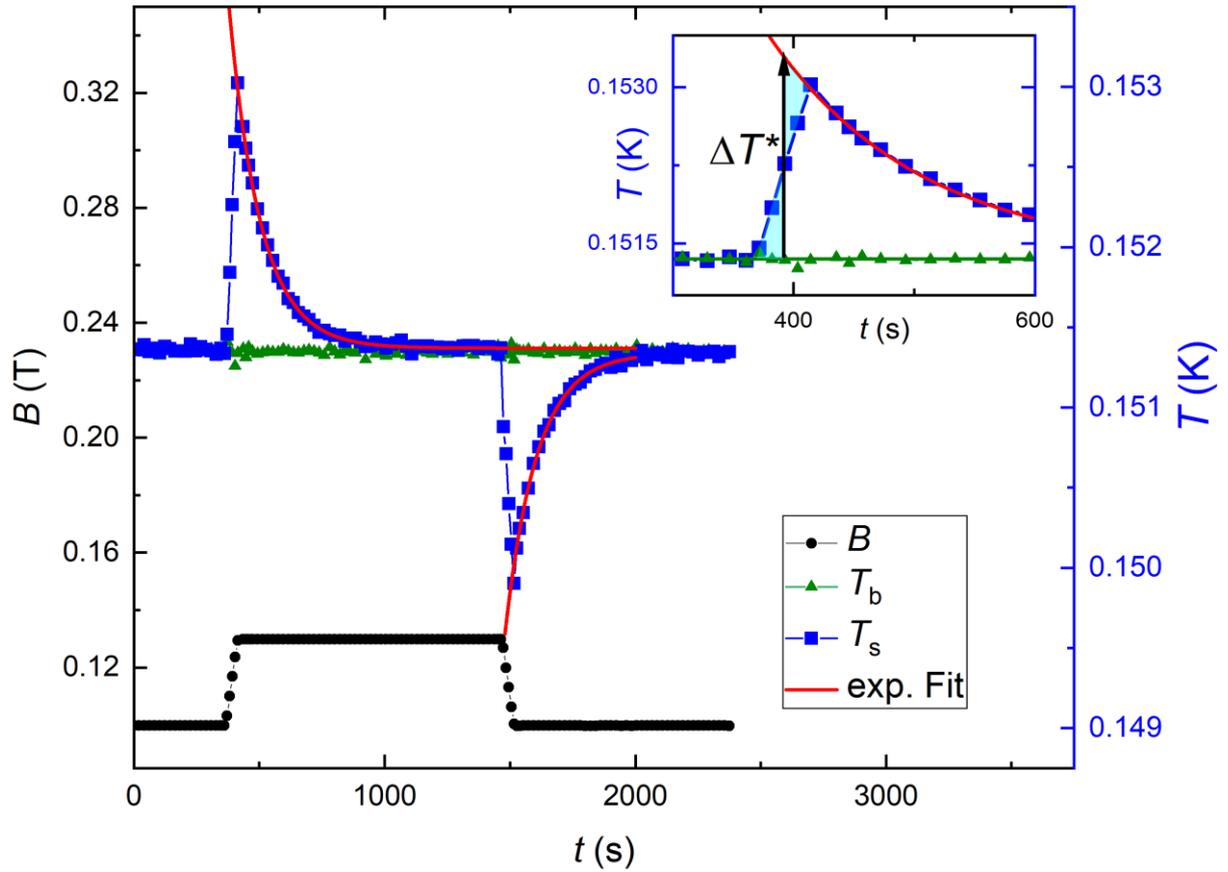

**Figure 3**. Illustration of a typical measuring cycle for determining the magnetic Grüneisen parameter. At the beginning of the cycle at time $t = 0$, the sample at the temperature $T_s$ is in thermal equilibrium with the bath temperature at $T_s = T_b \approx 0.151$ K. In the example shown, the sample is in a dc field of $B = 0.1$ T. At $t = 400$ s the field is ramped up by $\Delta B = +0.03$ T which is accompanied by an increase in $T_s$. After the field change is completed, $T_s$ relaxes back to $T_b$. This relaxation is fitted by an exponential decay $T_s(t) \propto \exp(-t/\tau)$ with $\tau$ the relaxation time (red solid line in Fig. 3). By extrapolating the exponential decay back to times at which the field change started, the increase in temperature $\Delta T^*$ can be determined from an equal-areas construction as indicated in the inset of Fig. 3.

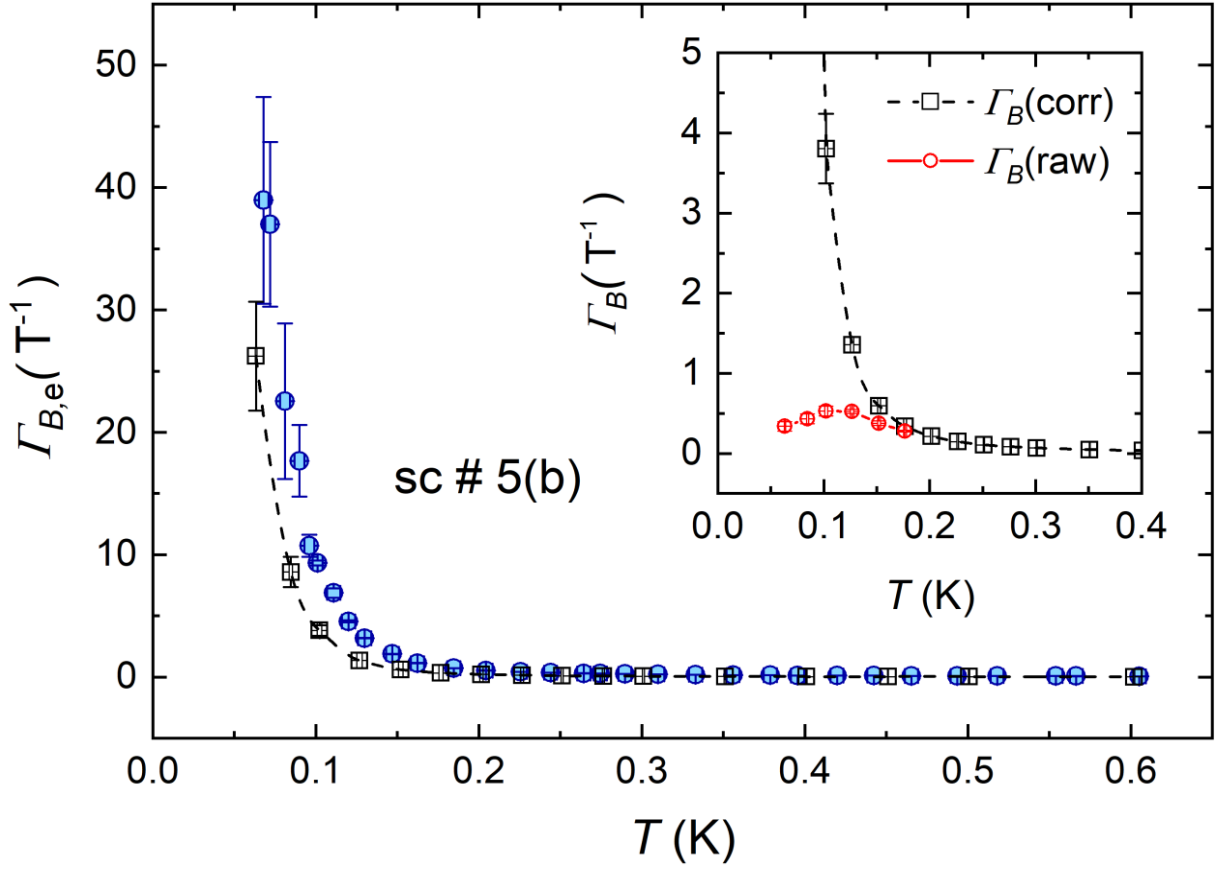

**Figure 4**. Temperature dependence of the electronic contribution $\Gamma_{B,e}$ to the magnetic Grüneisen parameter of single crystalline $PbCuTe_2O_6$. The main panel shows $\Gamma_{B,e}$ for sc #5(b), associated with the $Cu^{2+}$ spins, obtained by applying technique (1) described in the text (black open squares) and taking into account the nuclear contributions according to eq. (10). For comparison the figure also shows $\Gamma_{B,e}$ for sc #5(b) obtained by using technique (2) (blue spheres). The inset shows the raw data (red spheres), corresponding to $\Gamma_B$, and $\Gamma_{B,e}$ (black open squares) from the main panel for $T \leq 0.4$ K.

The magnetic Grüneisen parameter was also measured as a function of magnetic field at a constant temperature $T = 0.15$ K by using technique (1) described above. The results are shown in Fig. 5 as $1/\Gamma_{B,e}$ vs $B$. We find a rapid suppression following a $\Gamma_{B,e} \propto (B - B_c)^{-1}$ dependence with $B_c \approx 0$.

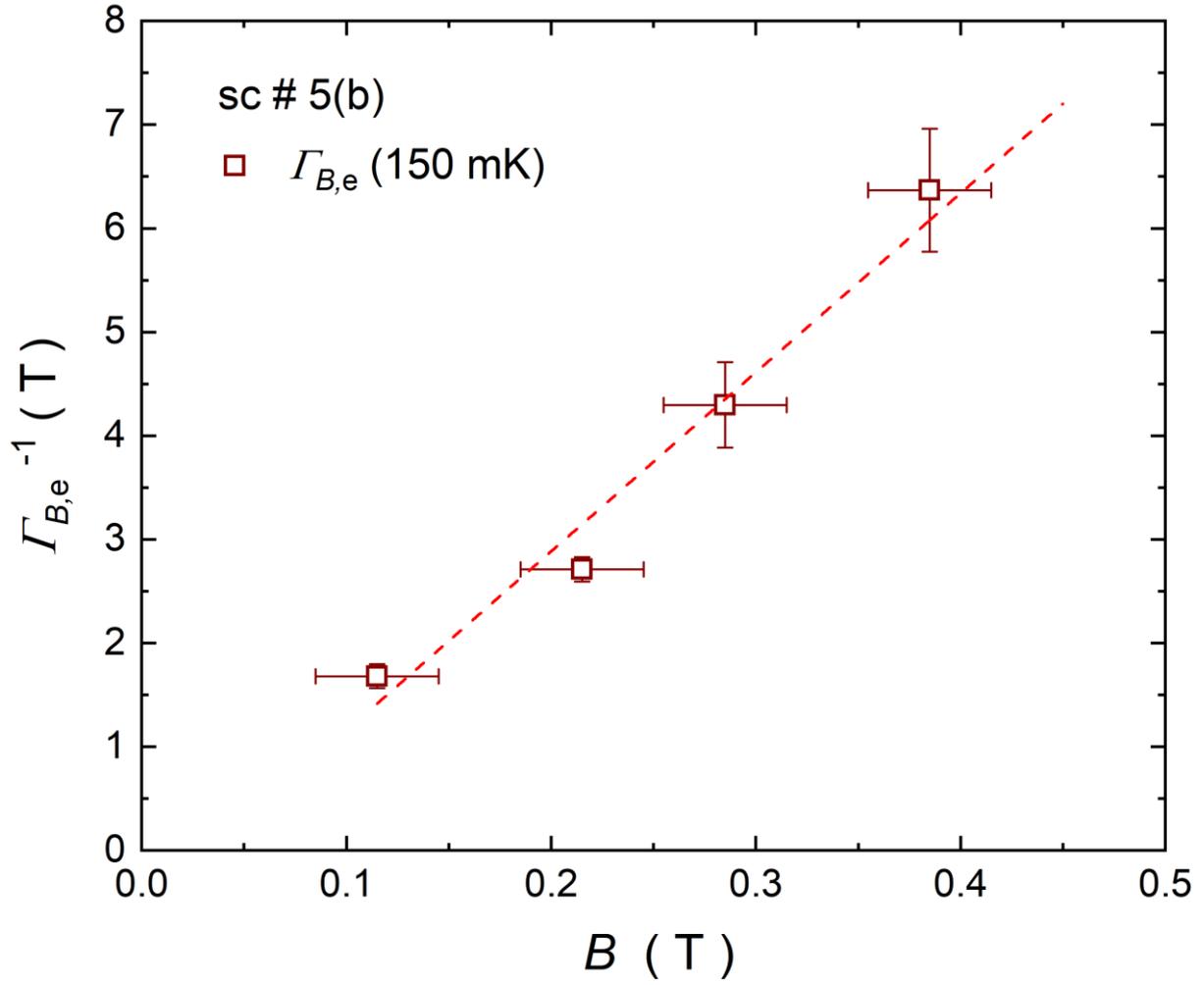

**Figure 5**. Electronic contribution to the magnetic Grüneisen parameter at a temperature $T$ = 0.15 K plotted as $1/\Gamma_{B,e}$ vs. $B$. The suppression of $\Gamma_{B,e}$ with field approximately follows a behaviour $\Gamma_{B,e}$ = - $A/(B - B_c)$ with $A$ = 0.058 ± 0.005 and $B_c$ = (0.033 ± 0.06) $T \approx 0$ (broken line).

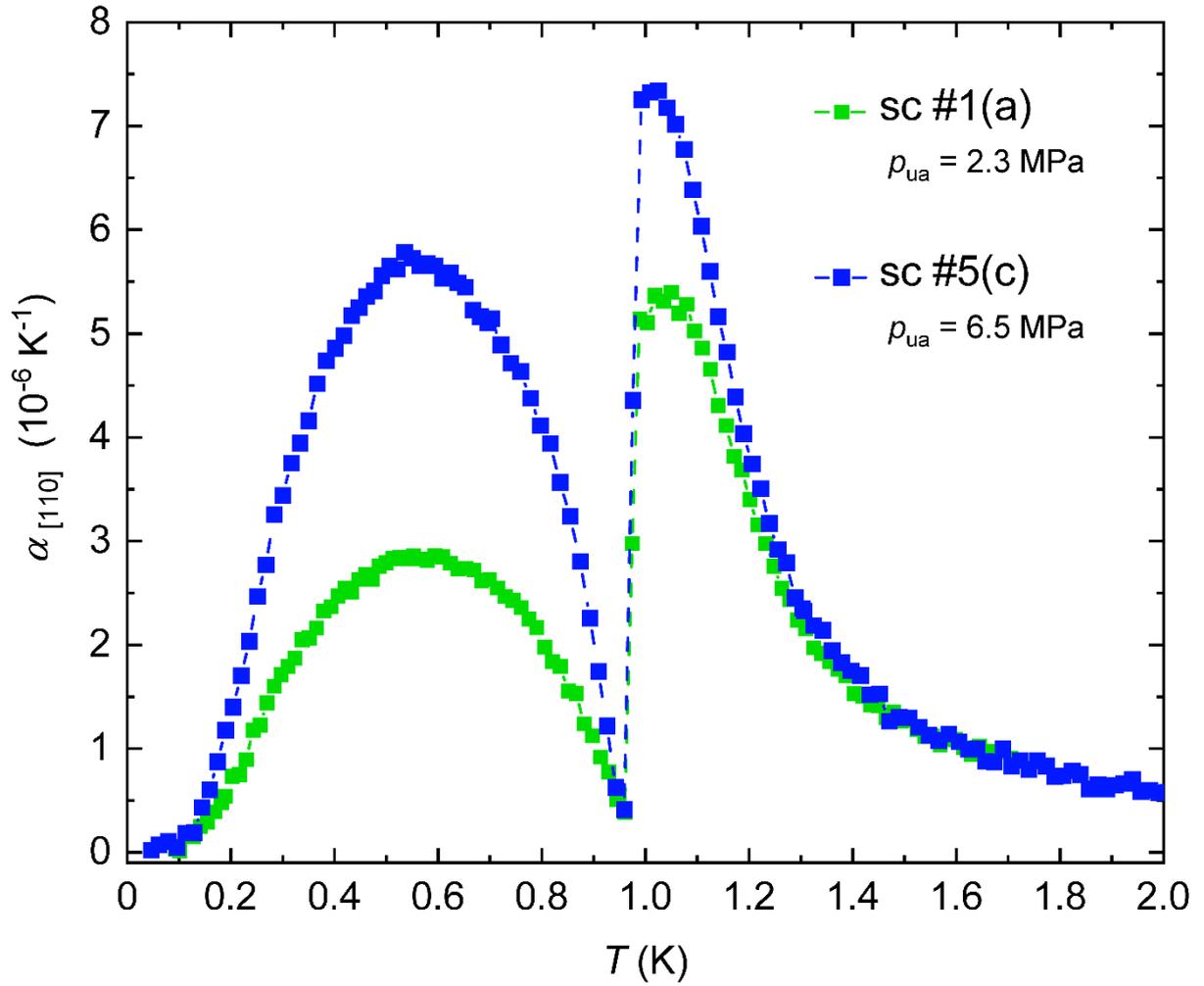

**Figure 6.** Coefficient of thermal expansion of PbCuTe$_2$O$_6$ single crystals sc #1(a) measured along the [1-10] direction and sc #5(c) measured along the [110] direction. In these measurements the uniaxial pressure along the measuring direction was $p_{ua}$ = (2.3 ± 0.5) MPa for sc #1(a) and $p_{ua}$ = (6.5 ± 1.3) MPa for sc #5(c). The pressure values refer to the pressure at around 4 K.

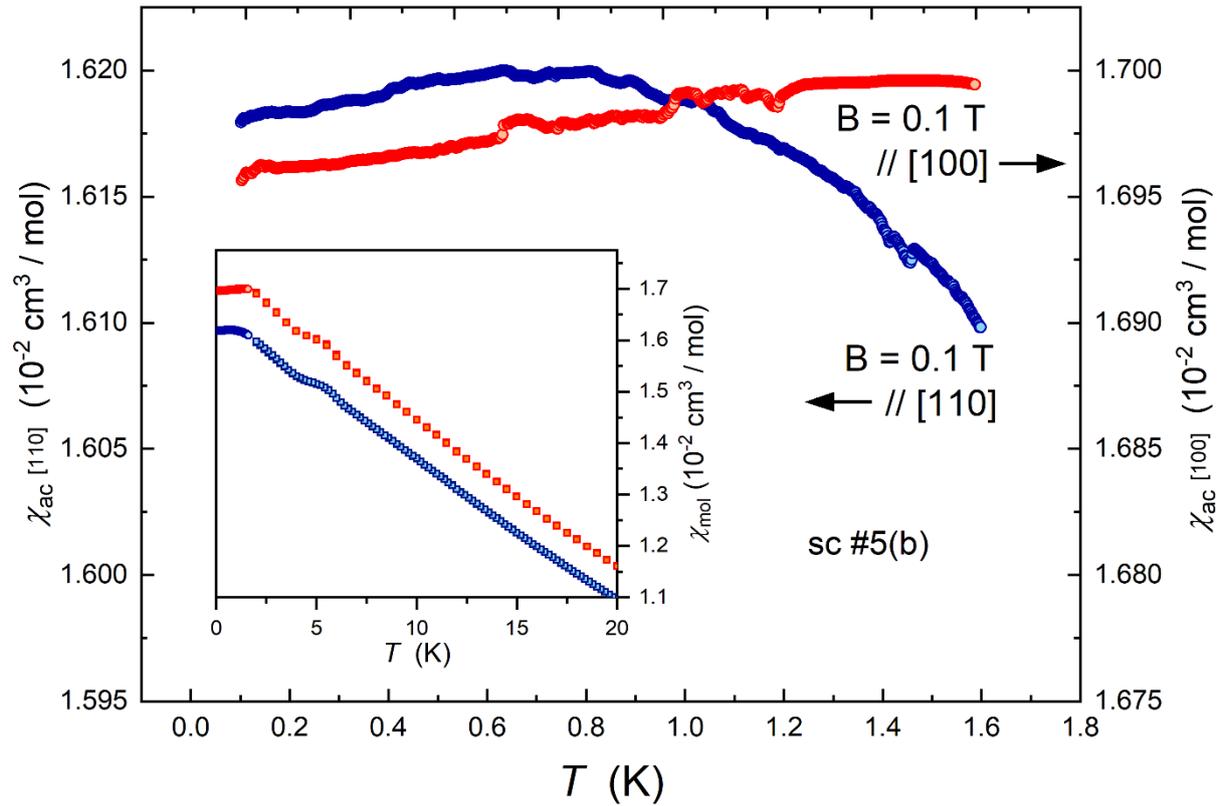

**Figure 7.** Temperature dependence of the ac-susceptibility, $\chi_{ac}$, of PbCuTe$_2$O$_6$ single crystal #5(b) for temperatures 0.1 K $\leq T \leq$ 1.6 K. The data were taken in a field of $B =$ 0.1 T applied along the [100] (red spheres, right scale) and [110] (blue spheres, left scale) directions. The inset shows the same data together with data (same colour code) for $T \geq 2$ K measured by using a commercial SQUID magnetometer (Quantum Design).

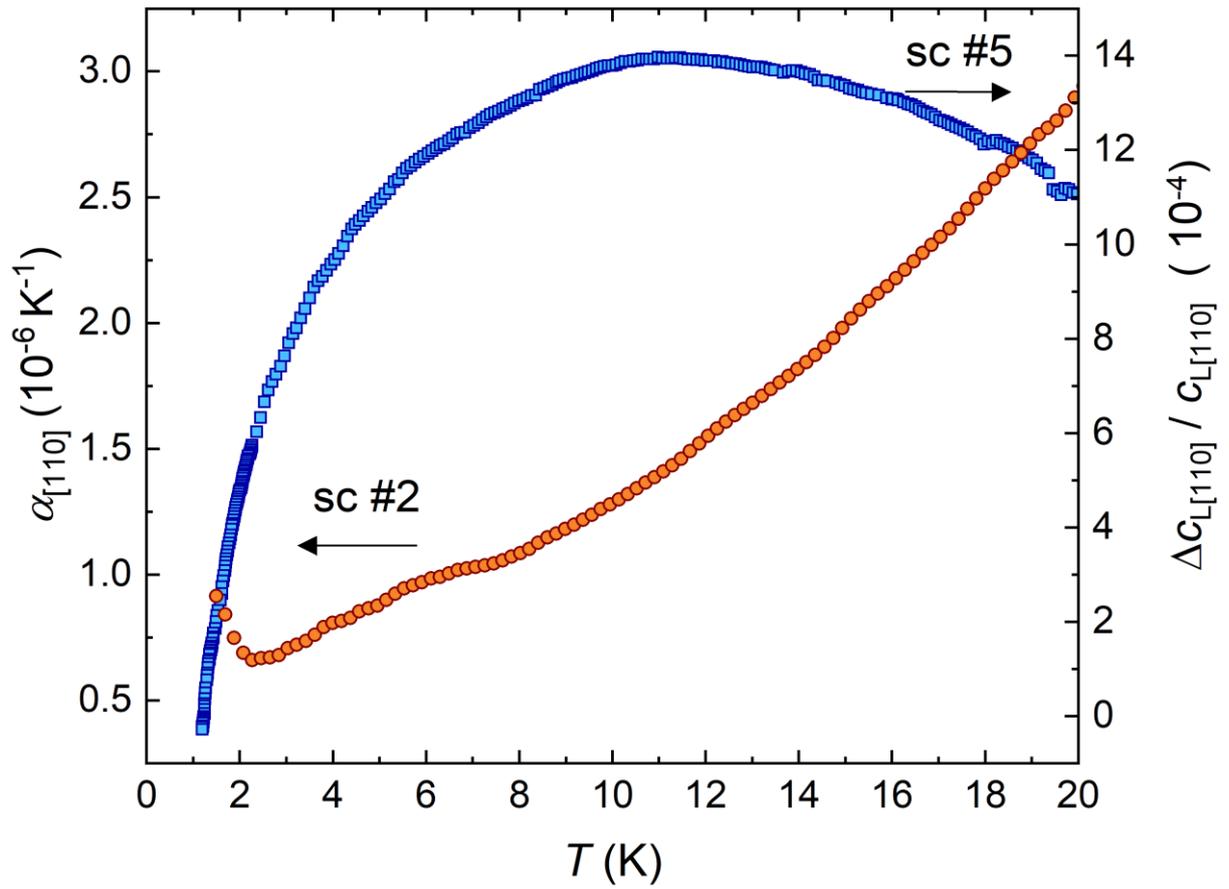

**Figure 8.** Coefficient of thermal expansion $\alpha_{[110]}$ for sc #2 together with the longitudinal elastic constant $c_{L[110]}$ for sc #5(a) for temperatures $1.3\ \text{K} \leq T \leq 20\ \text{K}$.

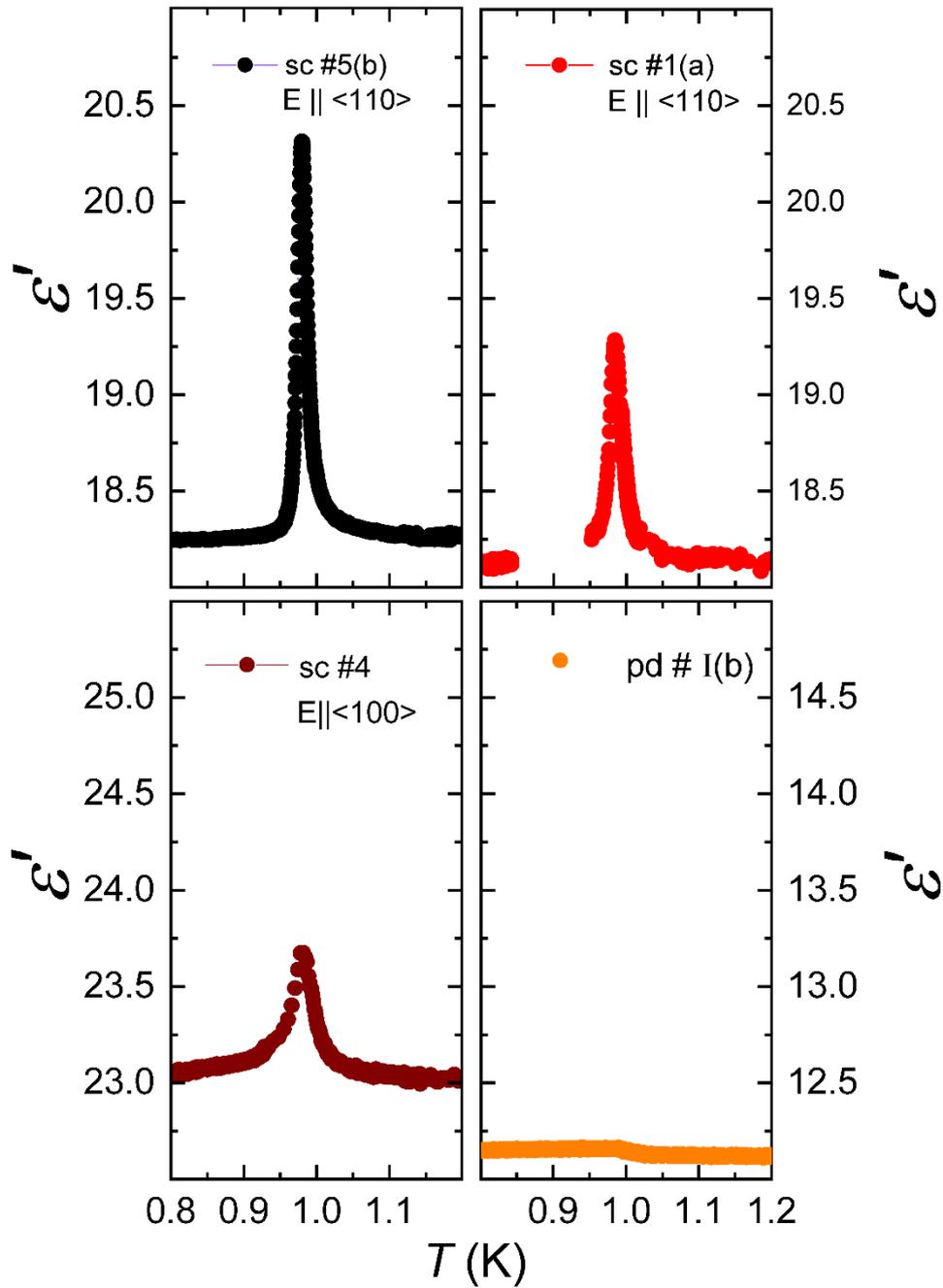

**Figure 9.** Results of the dielectric constant $\varepsilon'$ of PbCuTe$_2$O$_6$ for single crystals sc #5(b) (black spheres, upper left), sc #1(a) (light red spheres, upper right), sc #4 (dark red spheres, lower left), and a pressed-powder sample pd #I(b) (orange spheres, lower right). The orientation of the electric field **E** is indicated in the figure. Data were taken at a frequency of 11 kHz.

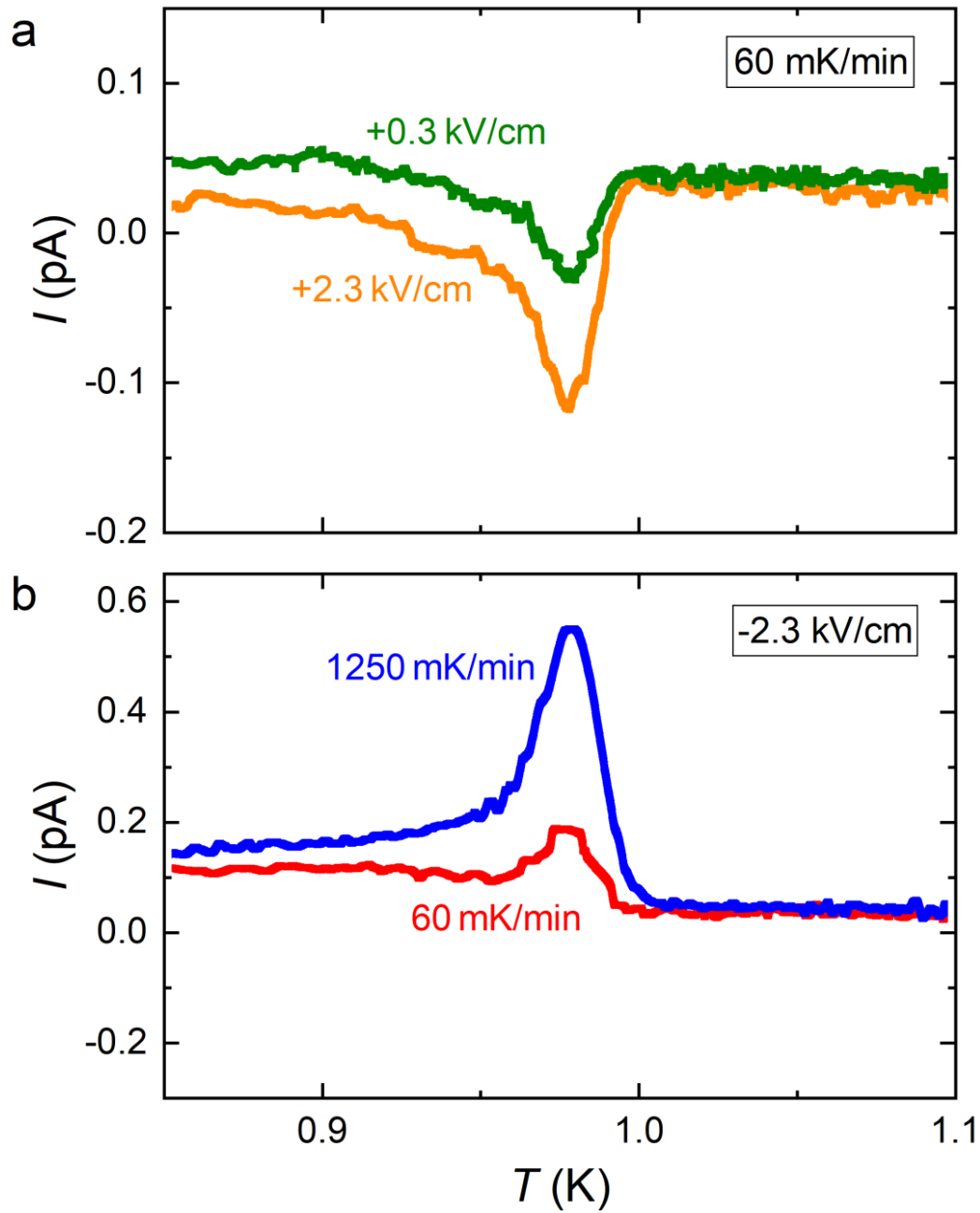

**Figure 10.** a: Pyrocurrent as measured with identical heating rates for two different poling fields. b: Pyrocurrent measured with two different heating rates for identical poling field.

## Supplementary Tables

| Isotope | z | I | $\mu$ ($\mu_N$) |
|---|---|---|---|
| $^{63}$Cu | 0.692 | 3/2 | 2.22 |
| $^{65}$Cu | 0.308 | 3/2 | 2.38 |
| $^{125}$Te | 0.143 | 1/2 | 0.887 |
| $^{207}$Pb | 0.221 | 1/2 | 0.578 |

$\mu_N = 5.05095 \cdot 10^{-27}$ J T$^{-1}$

**Table 1**. List of isotopes with their natural abundances $z$, their nuclear spins $I$ and the associated nuclear magnetic moment $\mu$ in units of the nuclear magneton $\mu_N$ taken from [4,5].

## Supplementary References